\documentclass[%
 reprint,
 amsmath,amssymb,
 aps,
]{revtex4-1}
\usepackage{graphicx,epstopdf}
\usepackage{dcolumn}
\usepackage{bm}
\usepackage{float} 
\usepackage{enumerate}
\usepackage[sort&compress]{natbib}
\usepackage{xcolor}
\usepackage{amsmath}
\usepackage{adjustbox}

\begin{document}


\title{Structural behaviour of supercritical fluids under confinement}

\author{Kanka Ghosh}
 \altaffiliation{kankaghosh@physics.iitm.ac.in.}
\author{C.V.Krishnamurthy}%
 \email{cvkm@iitm.ac.in}
\affiliation{%
Department of Physics, Indian Institute of Technology Madras, Chennai-600036, India  
  }%




\begin{abstract}
The existence of the Frenkel line in the supercritical regime of a Lennard-Jones fluid shown through molecular dynamics (MD) simulations initially and later corroborated by experiments on Argon opens up possibilities of understanding the structure and dynamics of supercritical fluids in general and of the Frenkel line in particular. The location of the Frenkel line, which demarcates two distinct physical states, liquidlike and gaslike within the supercritical regime, has been established through MD simulations of the velocity auto-correlation (VACF) and Radial distribution Function (RDF). We, in this article, explore the changes in the structural features of supercritical LJ-fluid under partial confinement using atomistic walls for the first time. The study is carried out across the Frenkel line through a series of MD simulations considering a set of thermodynamics states in the supercritical regime (P = $5000$ bar, $240$K $\leqslant$ T $\leqslant$ $1500$K) of Argon well above the critical point. Confinement is partial, with atomistic walls located normal to $z$ and extending to "infinity" along the $x$ and $y$ directions. In the "liquidlike" regime of the supercritical phase, particles are found to be distributed in distinct layers along the $z$-axis with layer spacing less than one atomic diameter and the lateral RDF showing amorphous-like structure for specific spacings (packing frustration),  and non amorphous-like structure for other spacings. Increasing the rigidity of the atomistic walls is found to lead to stronger layering and increased structural order. For confinement with reflective walls, layers are found to form with one atomic diameter spacing and the lateral RDF showing close-packed structure for the smaller confinements. Translational order parameter and excess entropy assessment confirms the ordering taking place for atomistic wall and reflective wall confinements. In the "gaslike" regime of the supercritical phase, particle distribution along the spacing and the lateral RDF exhibit features not significantly different from that due to normal gas regime. The heterogeneity across Frenkel line, found to be present both in bulk and confined
systems, might cause the breakdown of the universal scaling between structure and dynamics of fluids necessitating the determination of a unique relationship between them.
\begin{description}
\item[PACS numbers]
47.11.Mn, 05.20.Jj, 65.20.De
\end{description}
\end{abstract}

\maketitle


\section{\label{sec:level1}Introduction}
With its high density features like liquids, large diffusivity like gases and excellent dissolving power, supercritical fluids are playing a significant role in purification and extraction processes of various industries \cite{Kiran2000, McHardy1998, Wang2009}. Around $40$ years back in their seminal paper, M.E.Fisher and B.Widom discussed liquid and gaslike supercritical states by observing the decay behaviour of pair correlation function at large distances using linear continuum and lattice models and challenged the existing description of supercritical fluid as a single homogeneous phase like other states of matter \cite{Fisher1969}. Since then, many experimental studies had been executed to validate the heterogeneous nature of the supercritical fluids \cite{Nishikawa1995,Morita1997,Gorelli2006,Simeoni2010}. G.G.Simeoni et al. \cite{Simeoni2010} carried out inelastic X-ray scattering and molecular dynamics simulation to find out a demarcation line between two dynamically different regime ("liquidlike" and "gaslike") in supercritical fluid around critical point called Widom line. In a recent study, however, it has been found that this Widom line doesn't obey the corresponding states principle and the transition lines differ with different fluids \cite{Banuti2017}. Few years back, the discovery of Frenkel line in the phase diagram further adds to the current state of knowledge regarding heterogeneity of the supercritical state of a fluid. It indicates that there exists two distinct phases on either side of the Frenkel line: the "liquidlike" and the "gaslike" states at any arbitrary high temperature and pressures \cite{Yang2015, Brazhkin12, Brazhkin2013, Bolmatovnew2013}. Thus the universal and dynamic Frenkel line is qualitatively different from the Widom line, which unlike Frenkel line, exists near to the critical point only \cite{Brazhkin12}. The transition between these two regimes along the Frenkel line occurs when the liquid relaxation time ($\tau$) becomes nearly equal to the Debye vibration period ($\tau_{D}$), when the system becomes unable to support shear modes\cite{Yang2015}. Nevertheless, the more convenient approach to detect Frenkel line on the phase diagram from atomistic simulation is to monitor the disappearance of oscillations in Velocity auto-correlation function (VACF) as proposed by V.V.Brazhkin et al. \cite{Brazhkin2013}.\\
Structural and thermodynamic properties, associated with Frenkel line crossover, have been deduced from MD simulations in terms of RDF and specific heat capacity \cite{Dima2013}. D.Bolmatov et al. experimentally proved the presence of thermodynamic boundary associated with Frenkel line from a diffraction experiment on supercritical Argon in a diamond anvil cell (DAC) \cite{Bolmatov2015}. Extensive investigations have been done on the dynamic crossover of supercritical phases of water \cite{Fomin2015, Yang2015}, Iron \cite{Fomin2014}, $CO_2$ \cite{Yang2015, Bol2014, Fomin2}, Argon \cite{Bolmatovnew2013, Bolmatov2015}, $CH_4$ \cite{Yang2015} etc. 
\begin{figure}[H]
\centering
\includegraphics[width=0.5\textwidth]{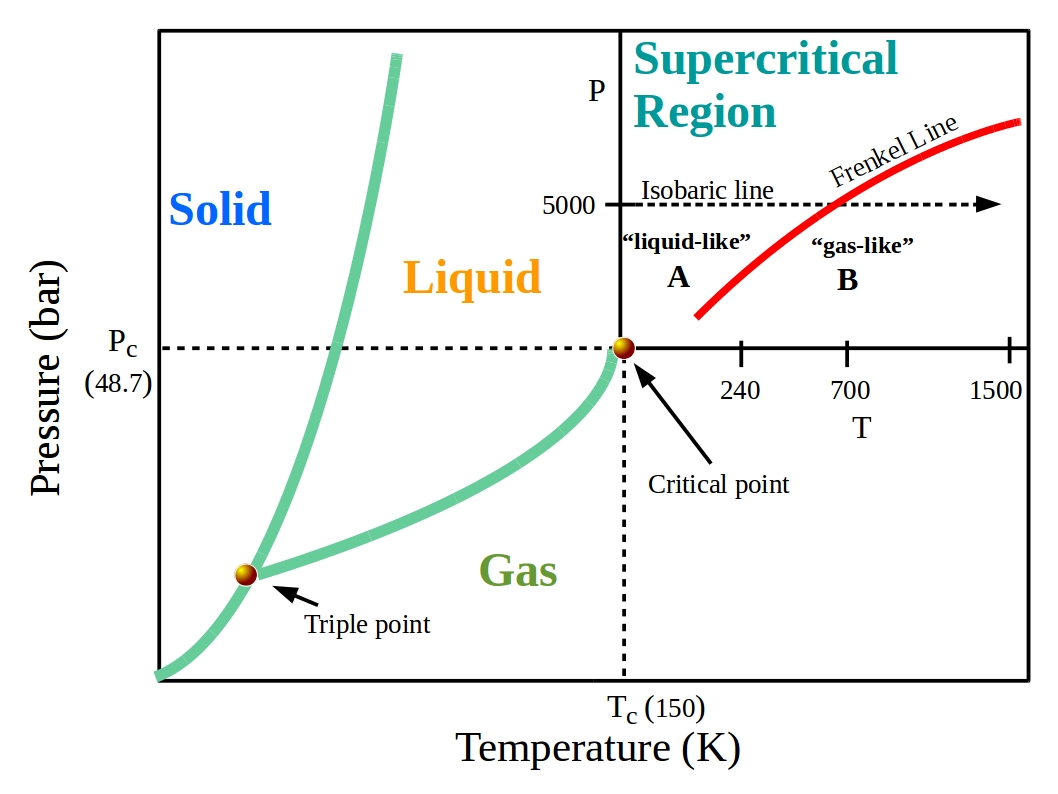}
\caption{\label{1} Phase diagram of the LJ fluid (Ar) in P-T plane. The isobaric line (P=$5000$ bar) of our study at supercritical phase is shown by the black dotted arrow.}
\end{figure}
In a recent review article, J.M.Stubbs covers a wide range of molecular simulation studies of supercritical fluids (SCF) \cite{Stubbs2016}. \\
Transport and structural behaviour of normal fluids under confinement has been of interest within the physics community due to their unusual properties with respect to the bulk fluid systems. J.Mittal et al. reported a series of studies on self-diffusion, modification of the dynamics and layering of confined hard-sphere fluids through Molecular Dynamics and Monte Carlo simulations \cite{Mittal2006, Mittal2008}. Recently, in a study of self-diffusion and radial distribution function of a strongly confined LJ fluid, N de Sousa et al.\cite{Sousa2016} found that in the solid-liquid phase transition region radial distribution function (RDF) corresponding to both the liquid and the solid phases are essentially indistinguishable.\\
In the present study, we choose atoms interacting with Lennard-Jones(LJ) potential mimicking Argon and consider a set of thermodynamics states in the supercritical regime (P = $5000$ bar, $240$ K $\leqslant$ T $\leqslant$ $1500$ K) of the bulk phase and determine the Frenkel line cross-over through VACF and RDF calculations to be at around T$\sim$ $600$-$700$K. We explore the changes in the structural features across the Frenkel line through a series of MD simulations of supercritical fluid under partial confinement using atomistic walls. Though the structural crossover in bulk supercritical fluid characterized by RDF peaks have been well established \cite{Dima2013}, the structural behaviour of confined supercritical fluids across the Frenkel line has not been studied yet. \\
We use atomistic boundaries on a pair of parallel sides along $z$ direction to simulate the partially confined systems. The simulation domain is taken to be a cuboid. Layering phenomenon under confinement has been observed and studied in detail before crossing , after crossing and in the close vicinity of the Frenkel line. The parallel and perpendicular components of the radial distribution function have been systematically studied for a wide range of confined spacings and structural ordering due to confinement has been understood through pair-excess entropy and translational order parameter calculations normal to the walls. Further, the differences in structural properties of supercritical fluid under both smooth, purely reflective and atomistic walls have been studied. The details of the MD simulation method have been presented in Section.II . Results are discussed in Section.III. Section.IV provides the summary and conclusions. The results of the confinement under purely reflective walls have been discussed in the appendix.

\section{\label{sec:level1}Model and Simulation Details}

We carry out molecular dynamics calculations on LJ fluid using LAMMPS software package \cite{Lammps}. To determine the Frenkel line and consistency checks, we model $10^5$ particles of LJ fluid fitted to Argon properties (LJ potential, $\frac{\epsilon}{k_B}$=$120$K, $\sigma$=$3.4$ $\AA$) in a number of  isothermal-isobaric (NPT) ensembles in bulk supercritical phase with periodic boundary conditions imposed along each of the three dimensions. We move on the P-T phase diagram of Argon \cite{Brazhkin2013} along an isobaric path with a constant pressure of $5000$ bar and temperature ranging from $240$ K to $1500$ K using a Nose Hoover thermostat and barostat (Fig.\ref{1}). 

We use a cut-off of $20$ $\AA$ (used previously for Ar in supercritical regime \cite{Bolmatov2015}) and shift the potential to make the potential and force continuous at the cut-off. In order to shift the potential to zero at the cut-off we have added $\Delta \epsilon$ = -$9.94\times10^{-7}$ev to the potential, which is numerically, too small to affect the critical point significantly. After an energy minimization, standard velocity-verlet algorithm with a time-step ($\Delta$t) of $0.0001$ picosecond (ps) has been used to equilibrate the system up to $50$ ps followed by a $200$ ps production run to calculate and analyse the properties of interest and to perform consistency checks.\\
In our MD simulations for different P,T state points ($240$K to $700$K at $5000$bar) the difference between MD and experimental density from NIST database \cite{NIST} is found to be less than $1$ $\%$. Our simulation range extends upto temperature $1500$ K, but the NIST data \cite{NIST} is available only upto $700$ K at $5000$ bar. However, the systems are quite well behaved for these state points (T$\geqslant$ $700$K) with stabilized energy, density and the velocity distributions follow the Maxwellian distribution, with the value of the standard deviation being consistent with the analytical value ($\sqrt{k_BT/m}$) over the entire temperature range.\\
Partially confined systems of LJ supercritical fluids are simulated in a cuboid with parallel walls facing each other normal to the $z$ axis at $z$ = $\pm H/2$, $H$ being the separation between the walls. We employ atomistic walls for introducing atomistic roughness to the  boundaries (Fig.\ref{2}).
\begin{widetext}
\center
\begin{figure}[H]
\includegraphics[width=1.0\textwidth]{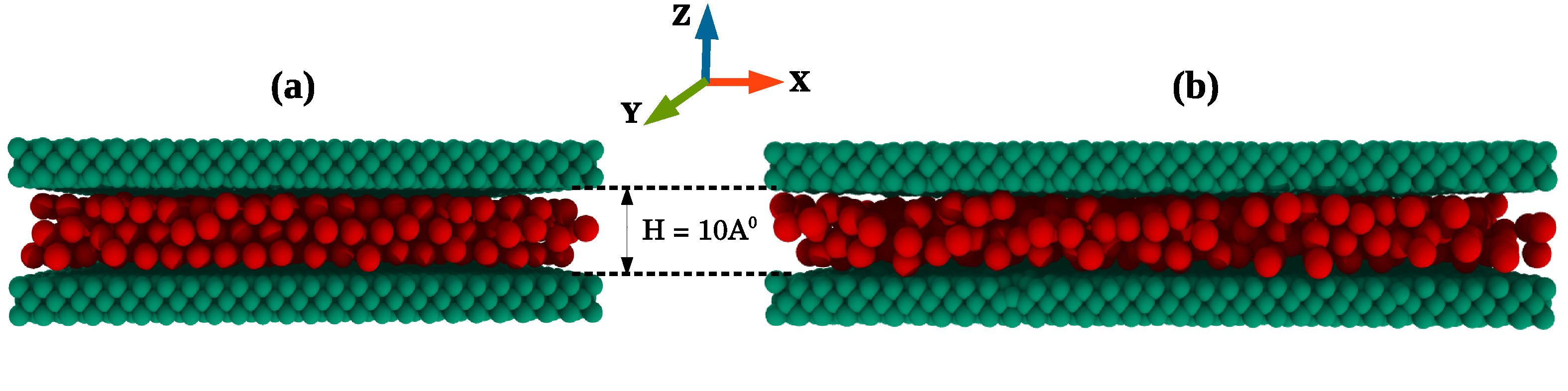}
\caption{\label{2} A snapshot of supercritical Argon confined between atomistic walls with a spacing H = $10$ $\AA$ between the walls at \textbf{(a)} $300$K (Before crossing the Frenkel line) and \textbf{(b)} $1500$K (After crossing the Frenkel line). Red atoms denote Argon and the light greenish-canary atoms represent wall-atoms(Ca). OVITO software is used to visualize the snapshot \cite{OVITO}.}
\end{figure}
\end{widetext}
For comparison we also simulate smooth, purely reflective walls at $z$ = $\pm H/2$. States with two different temperatures, $300$K and $1500$K at P = $5000$ bar are chosen. These two (P,T) state points in confinement lie on either side of the Frenkel line: while the point A($5000$ bar, $300$K) lies in the "liquidlike" regime, the point B($5000$ bar, $1500$K) lies in the "gaslike" regime. For each of these two (P,T) state points we have simulated different partially confined systems with different spacing between the walls which are varied from $1$ atomic diameter to about $21$ atomic diameters, keeping the density same as that of the bulk phase for the corresponding P,T point. Periodic boundary conditions are applied along both the $x$ and $y$ axes. To investigate the changes in structural behaviour of confined supercritical fluid (Ar) close to the Frenkel line we also consider a range of P,T points spanning both sides of the Frenkel line.\\
The solid, atomistic walls are made of $3$ layers of the face-centered cubic (fcc) lattice. The number of wall-atoms are varied from $1710076$ (H = $3.4$ $\AA$) to $83668$ (H = $70$ $\AA$) for the state point at $300$K and from $3164188$ (H = $3.4$ $\AA$) to $154588$ (H = $70$ $\AA$) for the state point at $1500$K. For modelling supercritical fluid confined between these two atomistic walls, $10^{5}$ Argon particles are simulated using NVT ensemble. The wall atoms are attached to the lattice sites by harmonic springs. We set the spring constant (k) for these springs as $1000$ $ev/ \AA^2$ for $300$K temperature to make the wall-atoms behave sufficiently rigid. We calculate the average mean squared displacement (MSD) of the wall-atoms and the average root-mean squared displacement (RMSD) of the wall atoms is found to be around $100$ times smaller than a typical atomic displacement of fluid particles ($\sim$ $1$ $\AA$) along confined direction ($z$). This confirms the sufficient rigidity of the walls. At $1500$K temperature, keeping same spring constant (k = $1000$ $ev$/$\AA^2$) the RMSD of the wall atoms is found to be around $40$ times smaller than a typical atomic displacement of fluid particles which assures moderate rigidity of the wall-atoms. To find out the implications of the rigidity of the walls on the structure, we also study the structural behaviour of supercritical Argon by varying the k-values for a specific confined spacing.\\
LJ potential has been used to model the interactions between both wall-fluid and wall-wall atoms. To model simple yet realistic walls, values of mass and the size of the wall-particles are taken from the calcium crystal data \cite{Markvoort2005}, where mass of the each wall-atom is taken as $40.078$ $a.m.u$. The LJ interaction parameters between the wall-atoms, $\epsilon_{wall-wall}$ = $0.2152$ev  and $\sigma_{wall}$ = $3.6$ $\AA$ are used. We use two different fluid(Ar)-wall LJ interaction strengths ($\epsilon_{wall-Ar}$) for our model: one same as the Ar-Ar interaction ($0.0103$ev) and the other a relatively stronger wall-Ar interaction ($\epsilon_{wall-Ar}$ = $0.0471$ev), obtained using Lorentz-Berthelot (LB) mixing rule ($\epsilon_{ij}$ = $\sqrt{\epsilon_{i} \epsilon_{j}}$ ) \citep{Allen1987}. The $\sigma_{wall-Ar}$ is taken as $3.5$ $\AA$ (LB-mixing rule: $\sigma_{ij}$ = ($\sigma_i$ + $\sigma_j$)/ $2$). The cut-off distance for fluid-wall interaction has been taken as purely repulsive type ($r_c:{wall-fluid}$ = $\sigma_{wall-Ar}$). The motion of wall-atoms is coupled to a thermostat of Nose-Hoover type to maintain same temperature as that of the supercritical fluid. This avoids any unnecessary heat flow though the fluid.  

\section{\label{sec:level1}Results and Discussions}
\subsection{\label{sec:level2} Identifying Frenkel line from VACF and RDF of bulk supercritical fluid}

The Velocity autocorrelation function (VACF) is generally defined as 
\begin{equation}
VACF(t)=\left\langle \vec{v}(0)\vec{v}(t) \right\rangle
\end{equation}
, where $\vec{v}(0)$ and $\vec{v}(t)$ are velocity vectors of particles at initial and some later time respectively and $\left\langle ... \right\rangle$ denotes the ensemble average. It is quite well understood that VACF is a monotonically decaying function for gases but it shows both oscillatory and decaying behaviour for liquids and solids \cite{Hansen2007, Brazhkin2013}. M.E.Fisher and B.Widom introduced long back the idea of using oscillatory and monotonous decay of pair-correlation function 
\begin{figure}[H]
\centering
\includegraphics[width=0.5\textwidth]{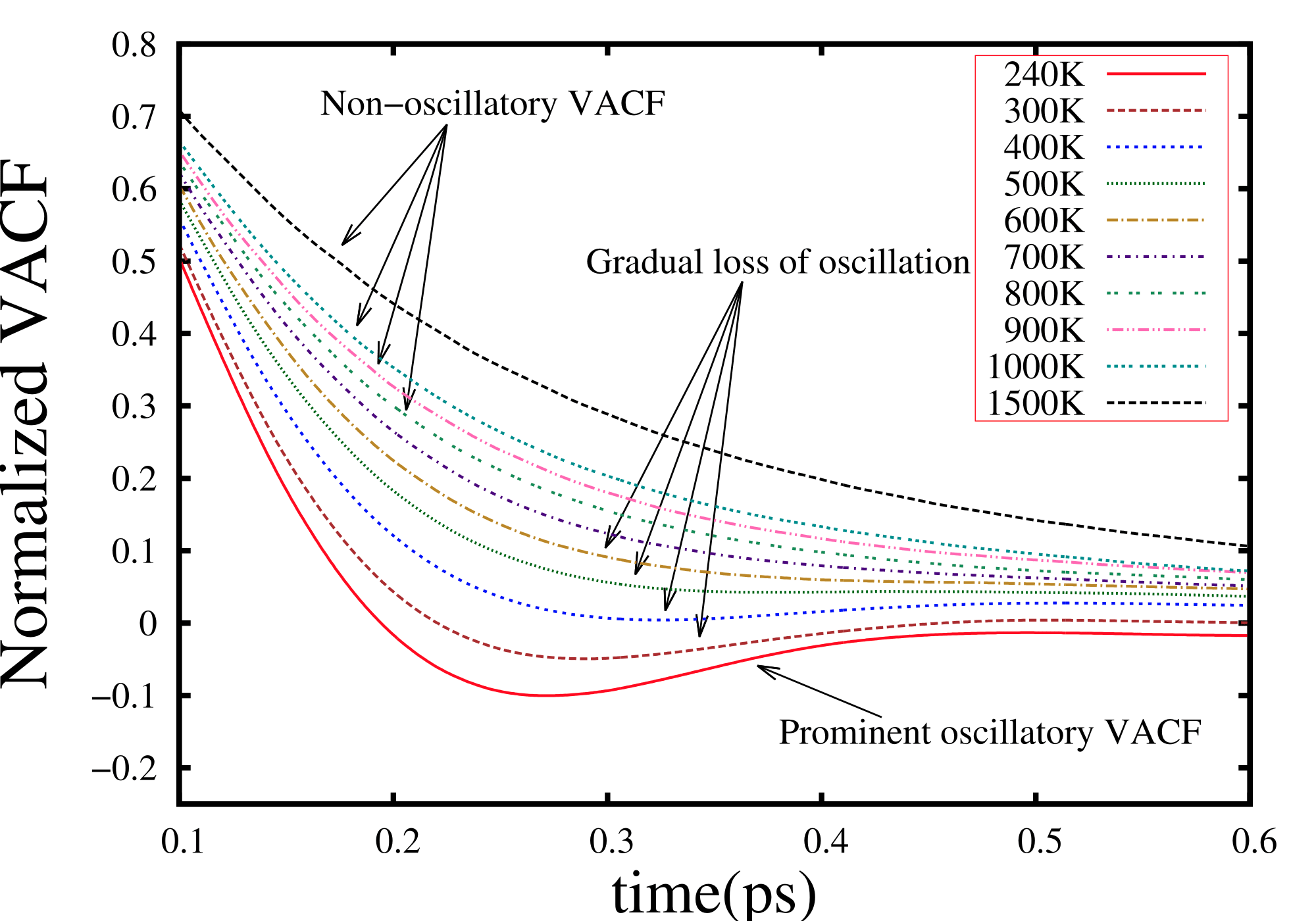}
\caption{\label{3} The Zoomed-in figure of the Velocity autocorrelation functions (VACF) of bulk-supercritical Argon for different temperatures ranging from $240$K to $1500$K along $5000$bar isobaric line in phase diagram. The VACF shows gradual decay of oscillations and around $600$-$700$K VACF becomes purely non-oscillatory ("gaslike").}
\end{figure}
as a marker to distinguish liquidlike and gaslike supercritical states using their analysis on linear continuum and lattice models \cite{Fisher1969}. Recently, V.V.Brazhkin et al. \cite{Brazhkin2013} showed that the minimum of VACF disappears when the supercritical fluid crosses the Frenkel line and goes from a "liquidlike" to a "gaslike" phase. We have chosen an isobaric line at $5000$ bar in the P-T phase diagram \cite{Brazhkin2013} of supercritical Argon (Fig.\ref{1}) and examined systems over a range of temperatures from $240$K to $1500$K. We observe the expected gradual loss of oscillation of the bulk VACF as we increase the temperature. Increasing the temperature helps the particles to overcome the transient cage like environment created by nearest neighbours and diffuse in a "gaslike" manner. We observe that this change to occur around T $\approx$ $600$-$700$K (Fig.\ref{3}).

\begin{figure}[H]
\centering
\includegraphics[width=0.5\textwidth]{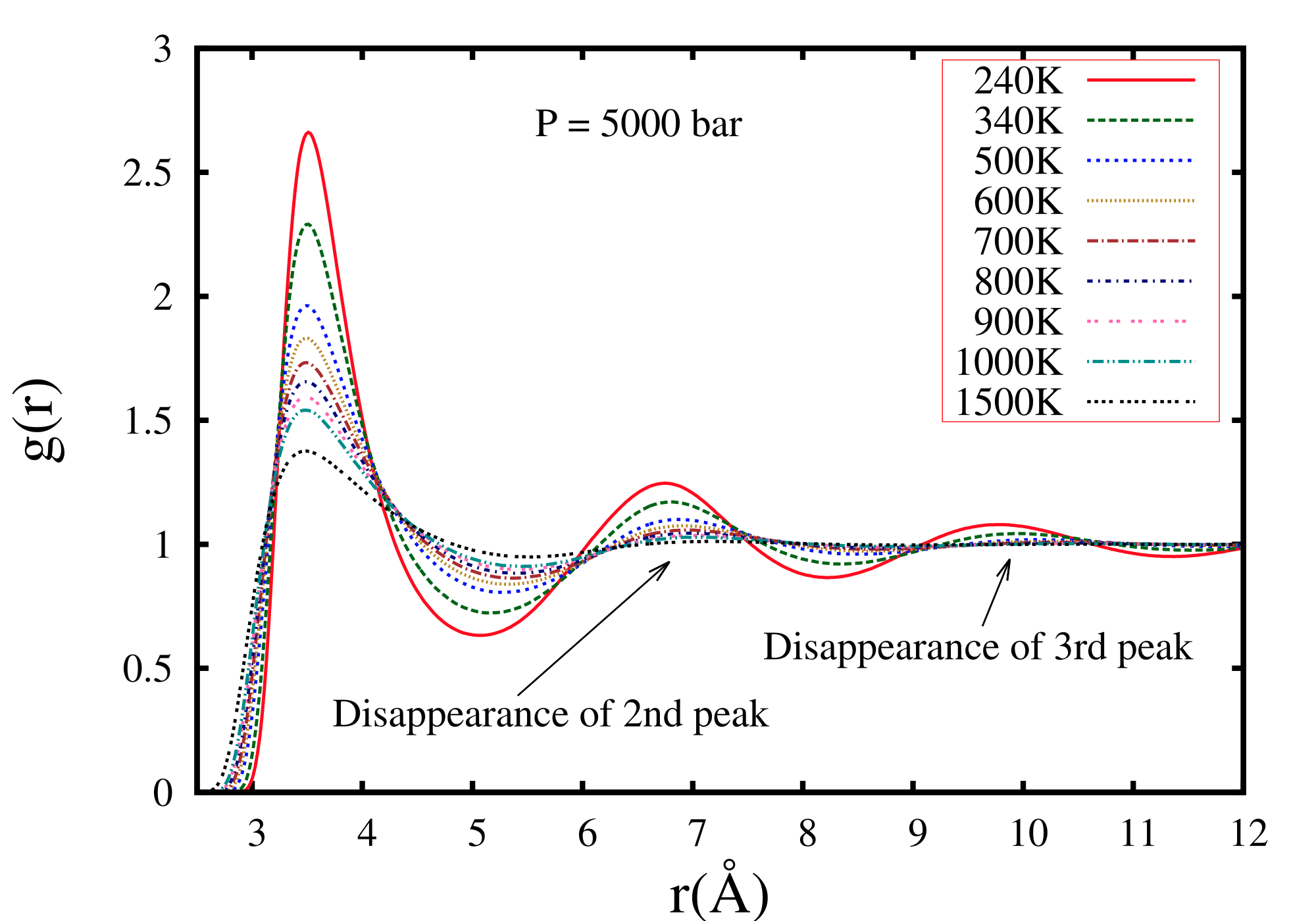}
\caption{\label{4}Zoomed-in figure of the structural crossover of the Radial distribution functions (RDF) of supercritical Argon for different temperatures ranging from $240$K to $1500$K along $5000$bar isobaric line in phase diagram. RDF makes gradual transition from a liquidlike to gaslike phase with a crossover at around $600$-$700$K.}
\end{figure}

Along with VACF, the structural properties of bulk SCF also undergo a well-defined change on crossing the Frenkel line as seen in the temperature evolution of the radial distribution function (RDF) at $5000$ bar pressure and depicted in Fig.\ref{4}. We observe the gradual decrement of the height of the first peak and near disappearance of $2$nd and $3$rd peak of bulk RDF after crossing the Frenkel line. The pronounced local ordering (giving rise to the oscillatory feature in VACF) gradually dies out after crossing the Frenkel line and the fluid undergoes a transition from a "rigid liquidlike" to "gaslike" state (Fig.\ref{4}). It is also found that the rate of decay of the first peak height is faster than the $2$nd peak height decay rate. It may be recalled that $g(r)$ is related to the coordination number, the total number of particles found within a given sphere, in the following way 
\begin{equation}
N(r) = \int_0^{r}g(r) 4\pi r^2 dr
\end{equation}
where N(r) is the number of particles found within the radius r. As the total number of particles is kept constant, the change in the local structure of the first coordination shell affects the structures in other neighbouring shells to satisfy the above equation. The integral implies that not all peaks can decay at the same rate with increasing temperature. While the first peak decays rapidly, the second peak decays at a relatively slower rate. We calculate the peak-heights and the first derivatives of the first and second peaks of RDF and vary them as a function of temperature. 

\begin{figure}[H]
\centering
\includegraphics[width=0.5\textwidth]{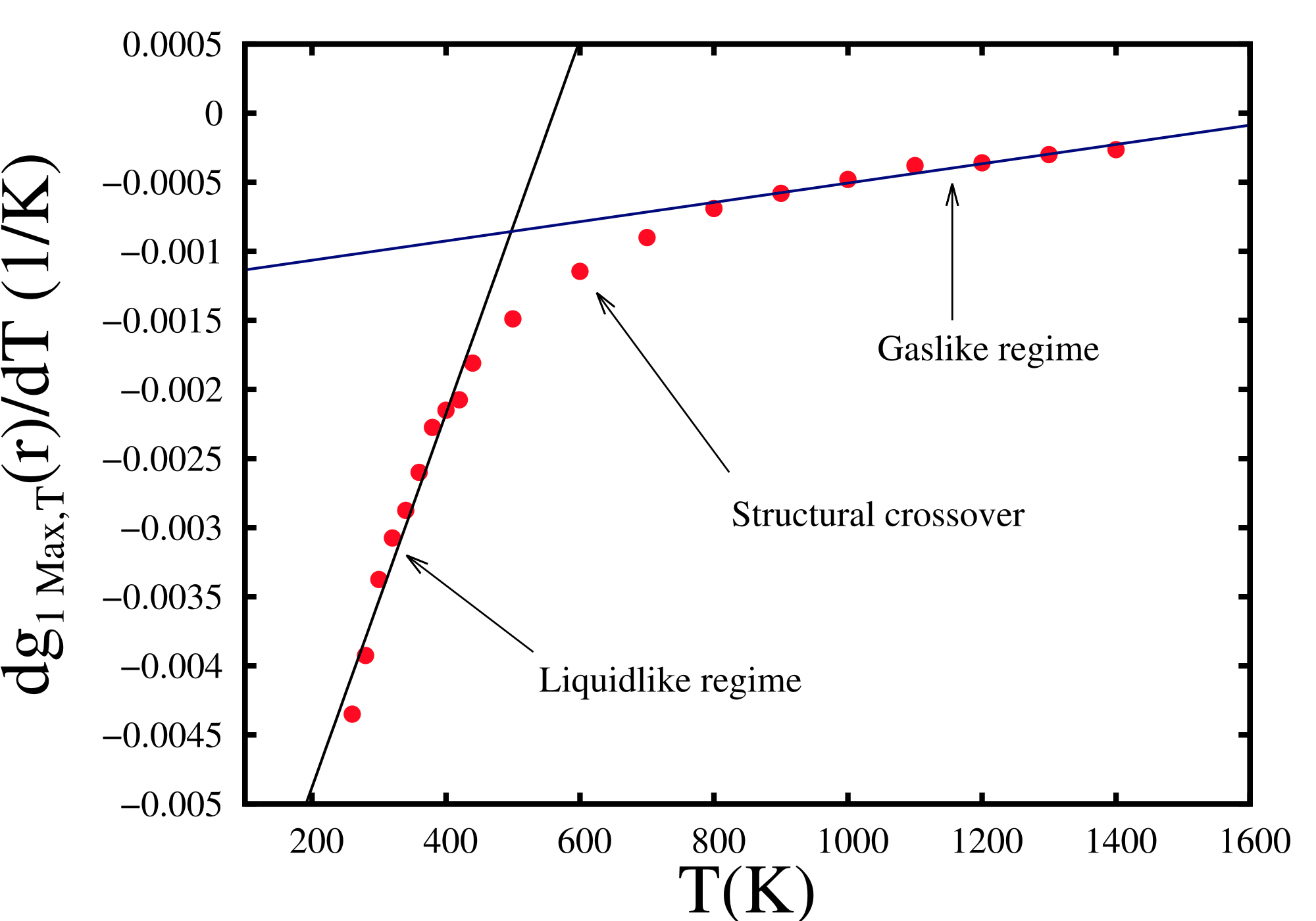}
\caption{\label{5}  Variation of the First order temperature derivative of the first peak-height of RDF with temperature (T).}
\end{figure}

\begin{figure}[H]
\centering
\includegraphics[width=0.5\textwidth]{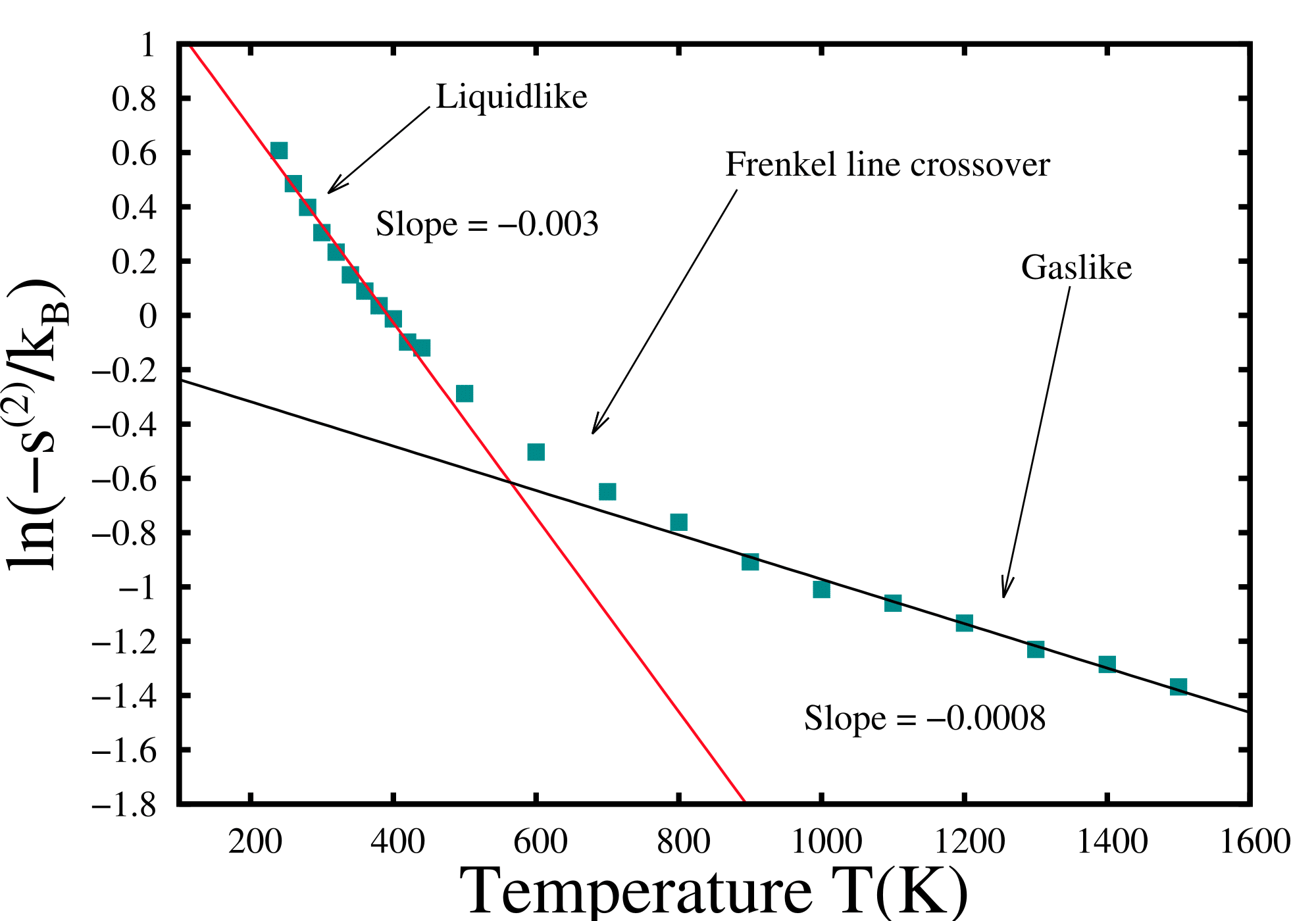}
\caption{\label{6} Scaled two-body excess entropy (ln($-s^{(2)}$ /$k_{B}$)) variation with temperature for bulk supercritical Argon. Liquid and gaslike regimes have distinct slopes and a crossover of structural ordering is evident along the Frenkel line (around $600$-$700$K) of supercritical Argon.}
\end{figure}

Two regimes can be noted from Fig.\ref{5}, which shows the derivative of the first peak-height as a function of temperature: one with a fast change (liquidlike regime) and the other with a slow change (gaslike regime)
of RDF peaks as a function of temperature. A structural crossover may be identified, by extrapolating the rates as shown in Fig.\ref{5}, to occur at around $600$-$700$K. The change across the Frenkel line can be seen to be gradual. Identical feature has been observed for the second peak-height of the RDF (see the appendix, Fig.\ref{23}). Similar observations have also been made by D.Bolmatov et al. \cite{Bolmatovnew2013}. \\
Two-body excess entropy is an alternate way of looking at how the degree of ordering changes with temperature. The presence of two distinct decay rates in RDF is manifested in the behaviour of the excess entropy as a function of temperature. This order parameter is defined \cite{Nettleton1958, Mountain1971} as
\begin{equation}
s^{(2)} = -2 \pi \rho k_B \int \left[g(r)lng(r)-g(r)+1\right]r^2dr
\end{equation}
where $\rho$ is the number density and $g(r)$ is the radial distribution function. It is well established that this two-body excess entropy ($s^{(2)}$) contributes approximately between $85$ $\%$ and $95$ $\%$ of the total excess entropy ($s_{ex}$)for a wide range of thermodynamic states for LJ fluids \cite{Baranyai1989}. Fig.\ref{6} shows the variation of the negative two body excess entropy ($-s^{(2)}/k_B$) in a logarithmic scale as a function of temperatures for bulk supercritical Argon. The gradual decay of $ln(-s^{(2)}/k_B)$ from a higher to lower values with increasing temperatures at a pressure of $5000$ bar indicates the gradual loss of ordering as we go from a low temperature(high density) to a high temperature(low density) regime.
Fig.\ref{6} shows two distinct regimes characterized by distinct slopes (-$0.003$ for liquidlike regime and -$0.0008$ for the gaslike regime). The crossover of ordering is found to occur at around $600$-$700$K temperature.\\
\subsection{\label{sec:level2} Number Density Fluctuations and Compressibility in bulk supercritical fluid}
We investigate another structural aspect, namely density fluctuations of the bulk supercritical LJ-fluid along the Frenkel line. It is well known that the number density fluctuations are closely related to the isothermal compressibility of the system through the relation
\begin{equation}
\kappa_{T} = \left(\frac{V}{k_{B}T}\right)\frac{\left\langle (\Delta \rho_{N})^2 \right\rangle}{\rho_{N}^2}
\end{equation} 
where, $\kappa_T$ is the isothermal compressibility, $\rho_N$ is number density ($\rho_N$ = $N/V$), $k_B$ is Boltzmann constant and $\left\langle...\right\rangle$ is the ensemble average \cite{McQuarrie1976}. The term $\frac{\left\langle (\Delta \rho_{N})^2 \right\rangle}{\rho_{N}^2}$ is nothing but the square of the standard deviation of the normalized number density ($\sigma_{\rho_N}^2$). We calculate the density fluctuations of supercritical Argon for each temperature ranging from $240$K to $1500$K in the form of distribution of densities. We find the $\sigma_{\rho_N}^2$ = $\frac{\left\langle (\Delta \rho_{N})^2 \right\rangle}{\rho_{N}^2}$ by calculating standard deviation of each of these distribution plots corresponding to each temperature. In Fig.\ref{7}.(a), we evaluate this temperature evolution of the density fluctuations for bulk supercritical Argon. We observe that at $5000$ bar the widths of the normalized number density fluctuations are increasing with increasing temperature. The values of corresponding standard deviations of the density distributions as well as isothermal compressibility ($\kappa_T$) calculated from the above formula of bulk supercritical Argon are given in tabular form in the appendix (Table.\ref{table:2}).\\

Thermodynamically, isothermal compressibility is defined as \cite{McQuarrie1976}
\begin{equation}
\kappa_T \equiv -\frac{1}{V}\left(\frac{\partial V}{\partial P}\right)_{N,T}
\end{equation}

It is quite well known that as a small change in pressure leads to a larger volume change in gases due to it's low density and packing with respect to liquids, the gases have higher compressibility than liquids. We found that the isothermal compressibility($\kappa_T$), evaluated from the standard deviation for the density fluctuations for bulk supercritical Argon, increases along the $5000$ bar isobaric line with increasing temperature from $240$K to $1500$K. In Fig.\ref{7}.(b) we have plotted $\kappa_T$ in logscale as a function of temperature, ranging from $240$K to $1500$K across the Frenkel line.
\onecolumngrid
\begin{widetext}
\begin{center}
\begin{figure}[H]
\begin{minipage}{0.5\textwidth}
\includegraphics[width=0.95\textwidth]{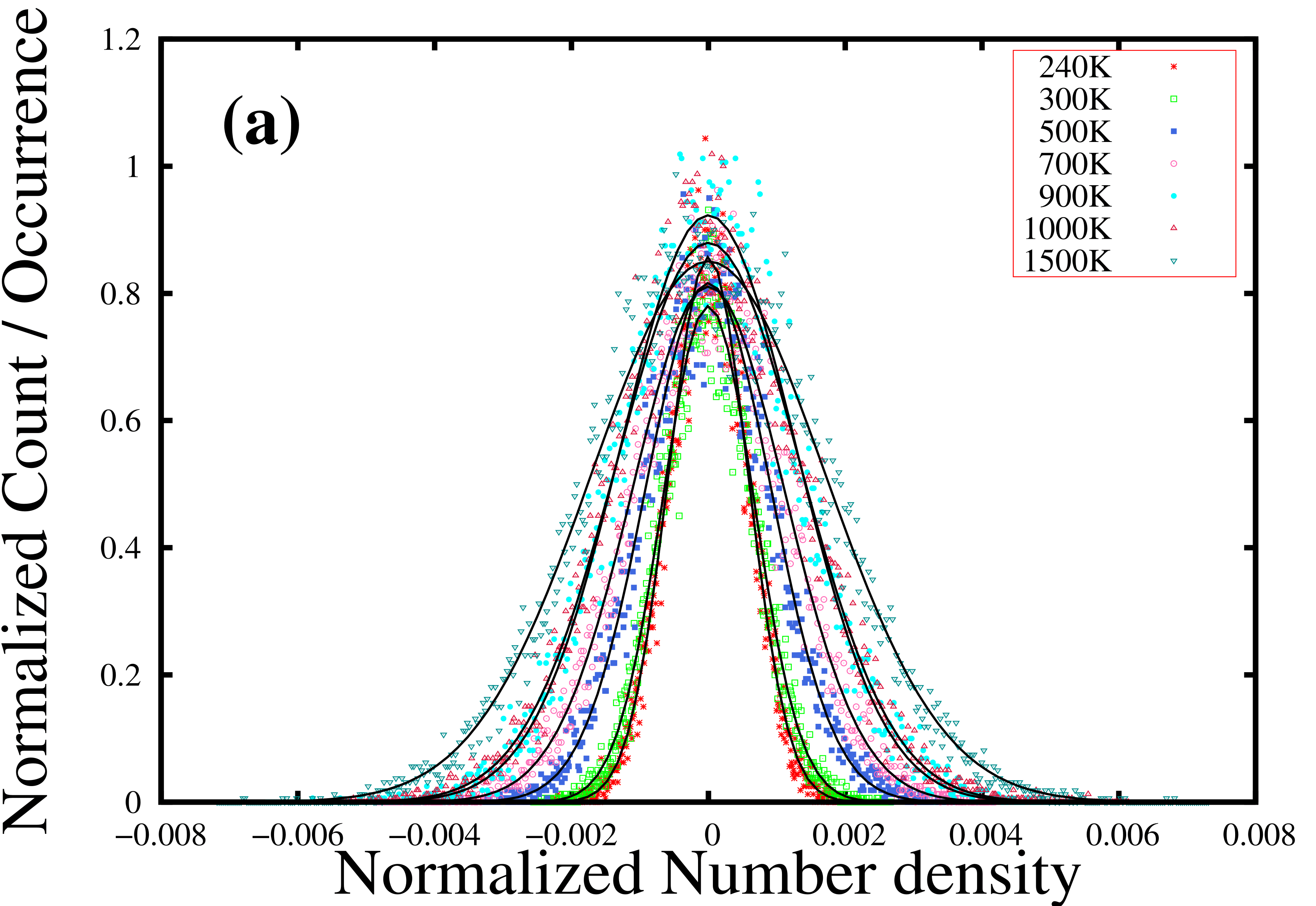}
\end{minipage}
\begin{minipage}{0.5\textwidth}
\includegraphics[width=0.95\textwidth]{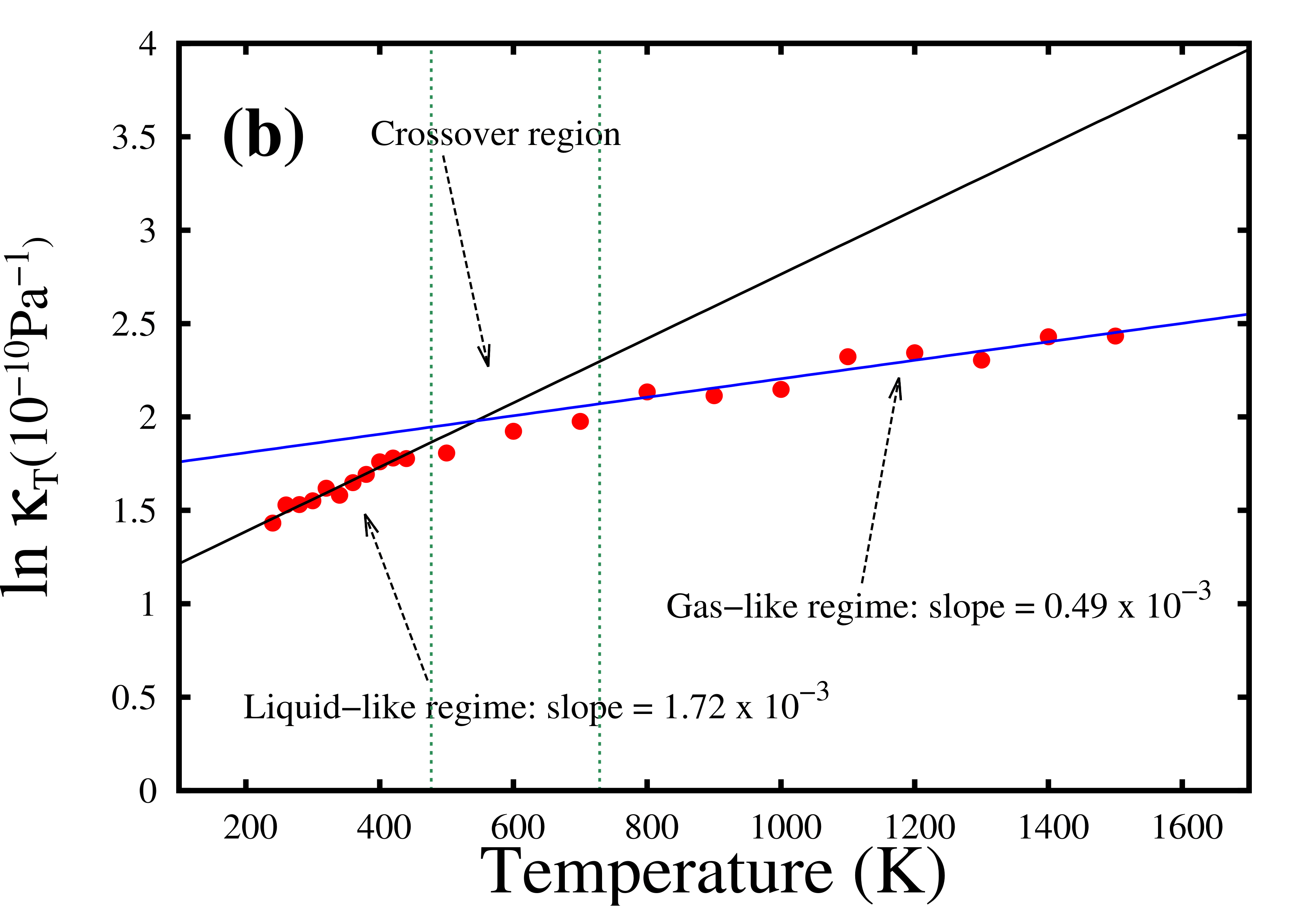}
\end{minipage}
\caption{\label{7} \textbf{(a)} The Number density fluctuations in the form of histogram for bulk supercritical Argon for different temperatures at $5000$ bar pressure along the Frenkel line. The best fit curves are shown in black. \textbf{(b)} Scaled isothermal compressibility (ln($\kappa_T$)) variation as a function of temperature (T). The variation of $\kappa_T$ with T has been found to behave in accordance with Frenkel line transition as two different straight lines with different slopes are needed to describe the entire range T dependency of $\kappa_T$ studied, with a crossover region in between($\approx$ $600$-$700$K).}
\end{figure}
\end{center}
\end{widetext}

We find that the slope of the fitted straight line changes as we go from lower to higher temperatures (at fixed pressure 5000 bar). The crossover or changing of the slope of the fitted line happens around $600$-$700$K temperature which can be interpreted as a Frenkel line transition of isothermal compressibility from a liquidlike to gaslike phase. Thus, the isothermal compressibility also undergoes a gradual crossover from a low compressible ("liquidlike") state to a highly compressible ("gaslike") state on crossing the Frenkel line.\\

\subsection{\label{sec:level2}Confinement studies of supercritical Argon along the Frenkel line}
To simulate partial confinement of supercritical LJ fluid, we employ atomistic LJ walls in a cuboid at $z$=$\pm H/2$, $H$ being the spacing between the walls and periodic boundary conditions along $x$ and $y$ directions. We investigate structural features of the system by considering wall separations in the range $3.4$ $\AA$ $\leqslant$ H $\leqslant$ $70$ $\AA$. 

\subsubsection{\label{sec:level3}\textbf{Distribution of LJ particles in supercritical phase normal to the walls :}}
As a first signature of the structural behaviour of the LJ fluid in supercritical phase under partial confinement, we investigate the distribution of the particles normal to the walls. Fig.\ref{8} shows some general features of the particle distribution normal to the walls for different wall-spacings at each of the two P,T state points A (P =$5000$ bar, T =$300$K) and B (P = $5000$ bar, T =$1500$K). Thus A represents "liquidlike" phase before crossing the Frenkel line and B represents "gaslike" phase after crossing the Frenkel line. Fig.\ref{8}.(a) shows particle distributions in phase "A" exhibiting distinct layering normal to the walls. For all the spacings considered, there is a depleted region close to the walls. For larger wall spacings the layering is seen to be prominent only near the walls, with the featureless central region describing the average bulk density. In sharp contrast, layering is absent in phase "B" as can be seen from fig.\ref{8}.(b). However, depletion region can be seen for all the spacings. In both phases ("A" and "B") the depletion region arises from repulsive interactions between the fluid and wall particles. The extent of the depletion region in all cases is of the order of $\frac{1}{2}$ atomic diameter at both ends. As the total number of particles is conserved, the particle number distribution exhibits (i) strong oscillatory features in the liquidlike high-density phase "A", (ii) weak oscillatory features in the gaslike phase "B".
\begin{widetext}
\begin{center}
\begin{figure}[H]
\hspace{1.5cm}\includegraphics[width=0.8\textwidth]{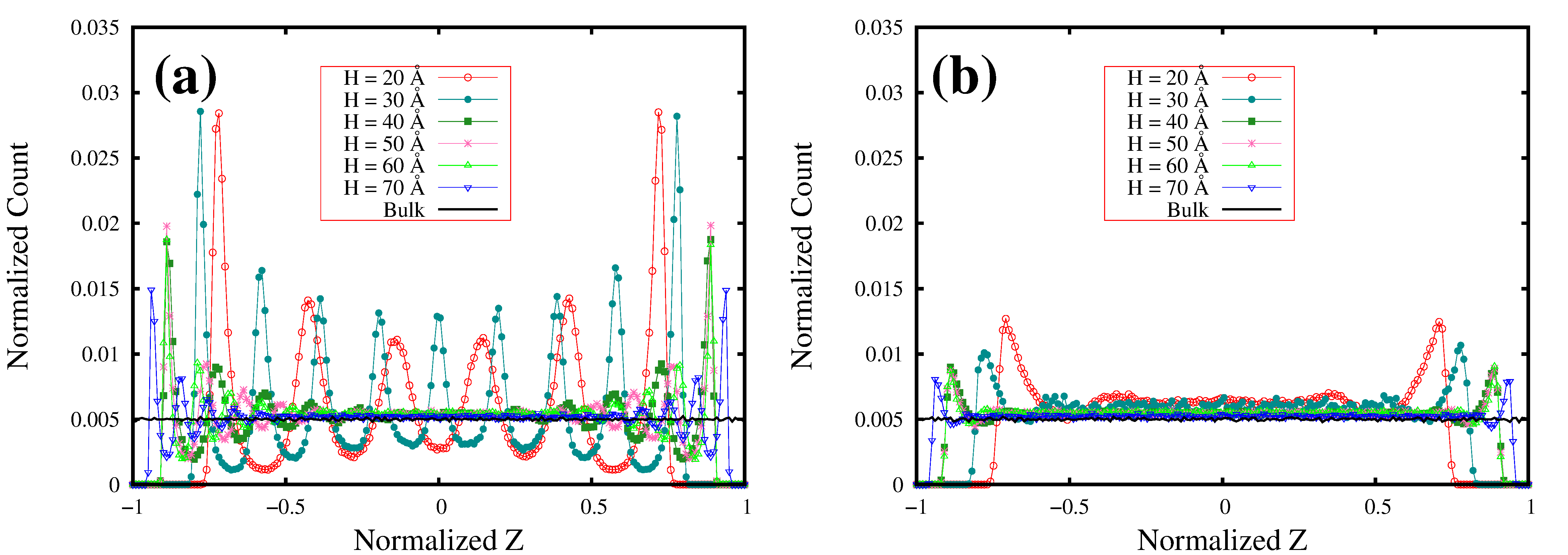}
\caption{\label{8} Distribution of Argon particles in supercritical phase under confinement at \textbf{(a)} $300$K (before crossing the Frenkel line) and at \textbf{(b)} $1500$K (after crossing the Frenkel line).}
\end{figure}
\end{center}
\end{widetext}
\subsubsection{\label{sec:level3}\textbf{The Structural features normal to the walls : Before crossing the Frenkel line}}
\begin{figure}[H]
\centering
\includegraphics[width=0.5\textwidth]{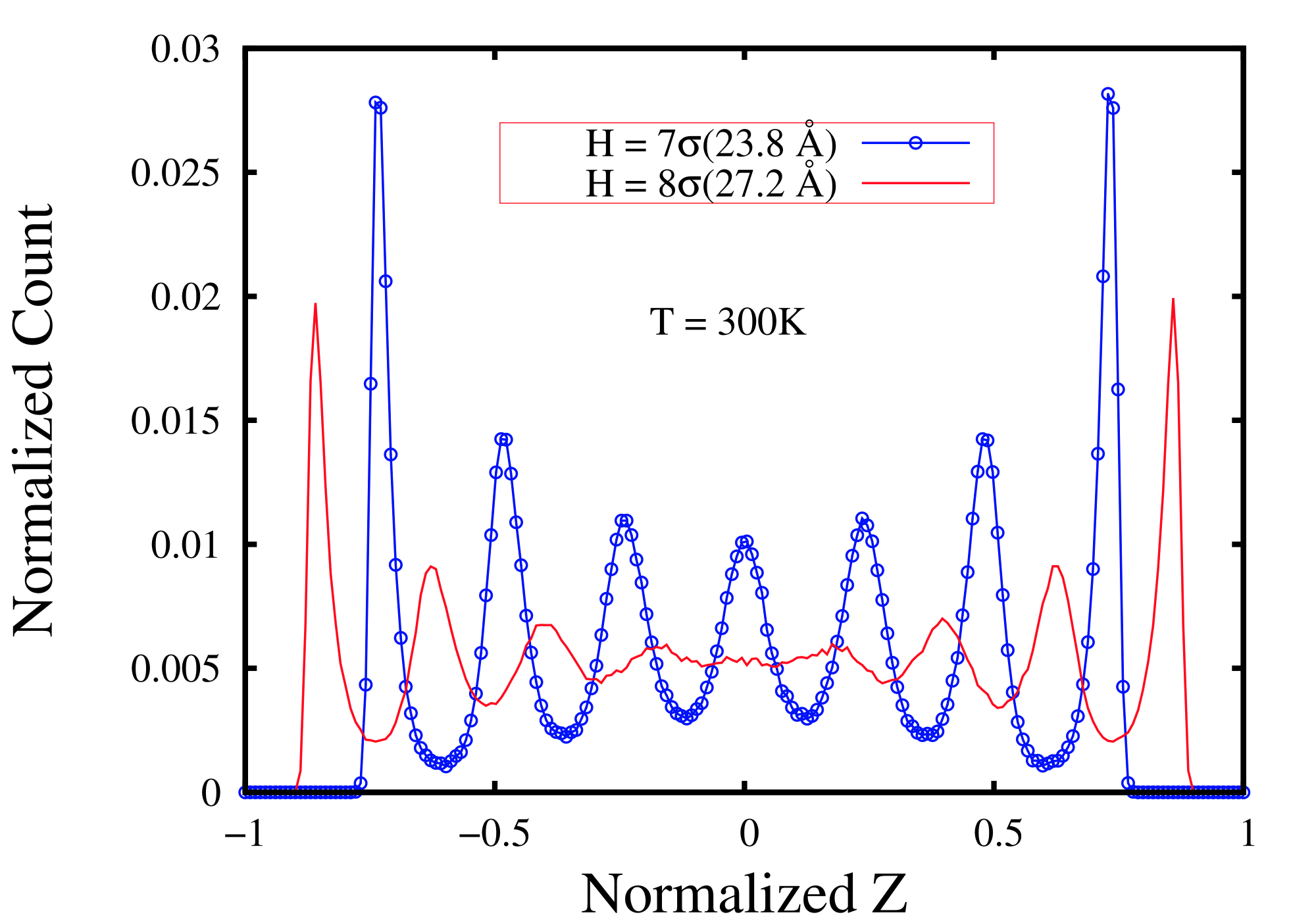}
\caption{\label{10} Normalized particle distribution along $z$ for two spacings: H = $23.8$ $\AA$ (= $7$$\sigma$) and H = $27.2$ $\AA$ (=$8$$\sigma$). While, H =$7$$\sigma$ is showing well formed peaks in the particle distribution profile, H = $8$$\sigma$ can't support well-developed layers normal to the walls, due to more packing frustration.}
\end{figure}

\begin{figure}[H]
\centering
\includegraphics[width=0.5\textwidth]{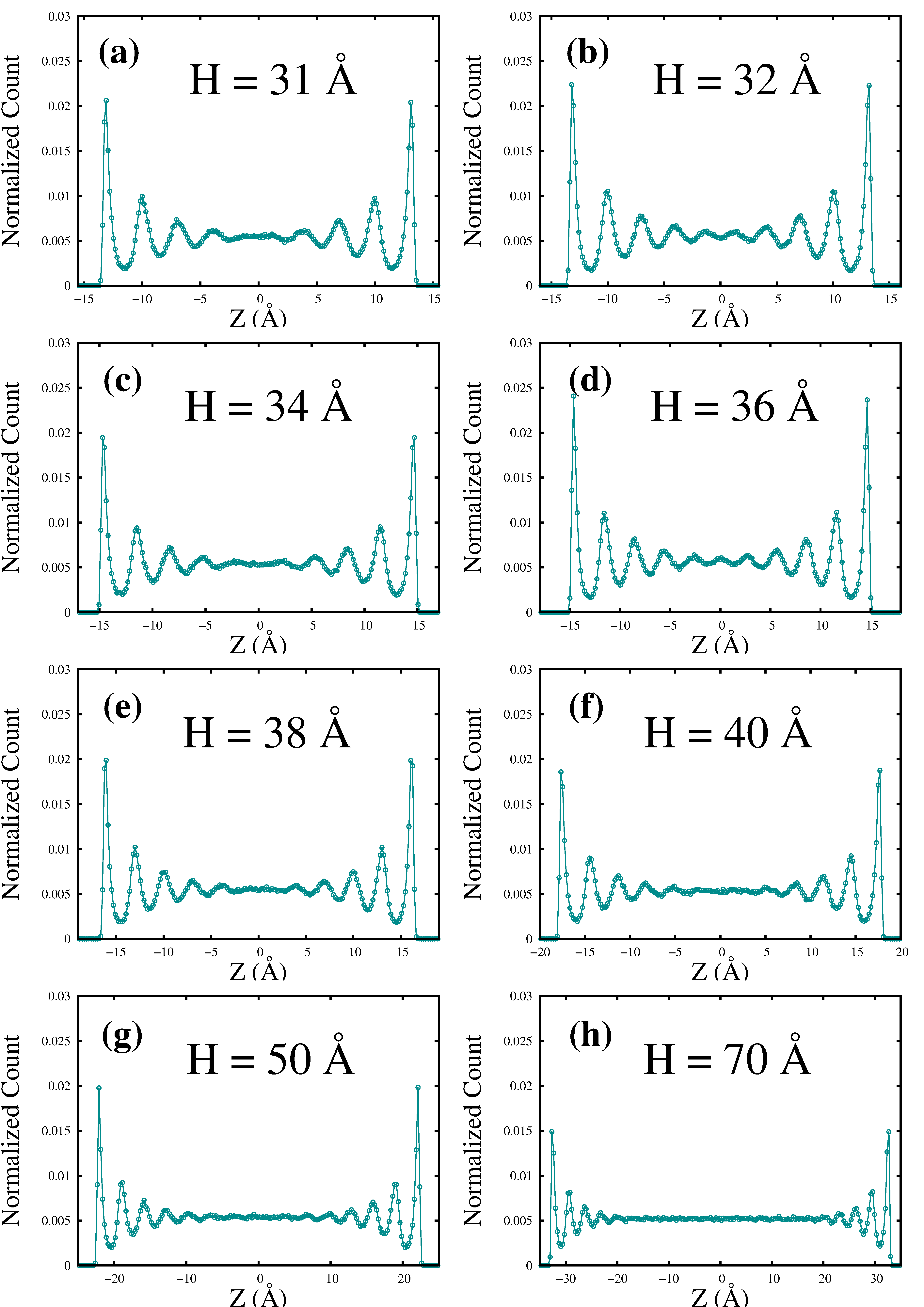}
\caption{\label{11} Distribution of Argon particles in supercritical state, averaged over several timesteps at $300$K normal to the walls for confined spacings(H $>$ $30$ $\AA$). \textbf{(a)} H = $31$ $\AA$, \textbf{(b)} H = $32$ $\AA$, \textbf{(c)} H = $34$ $\AA$, \textbf{(d)} H = $36$ $\AA$, \textbf{(e)} H = $38$ $\AA$, \textbf{(f)} H = $40$ $\AA$, \textbf{(g)} H = $50$ $\AA$, \textbf{(h)} H = $70$ $\AA$.}
\end{figure}

It is intriguing to note that in phase "A", the distance between two successive peaks in the number distribution profile is found to be, on average, lower than one atomic diameter of the particle. It implies that the particles are not arranged in  monolayers along the width of the spacings. It is believed to arise from the packing frustration (a competition between the preferred packing of the fluid particles and the packing constraints imposed by the confining walls \cite{Kim2014}) among the particles in supercritical phase. The effect of packing frustration may be seen more prominently for relatively smaller spacings. 
\begin{widetext}
\begin{center}
\begin{figure}[H]
\begin{minipage}{.5\textwidth}
\includegraphics[width=0.98\textwidth]{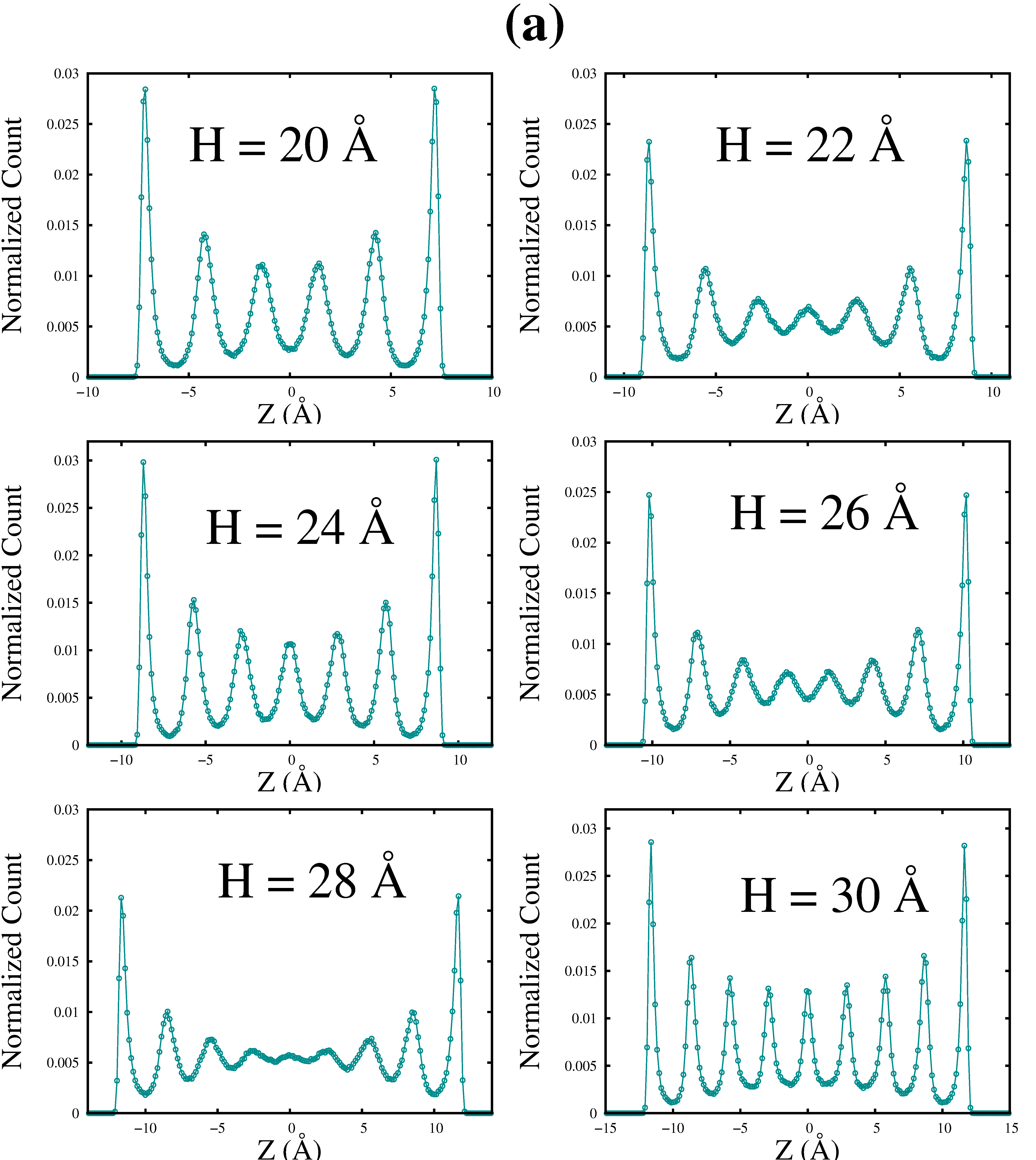}
\end{minipage}
\begin{minipage}{.5\textwidth}
\includegraphics[width=0.98\textwidth]{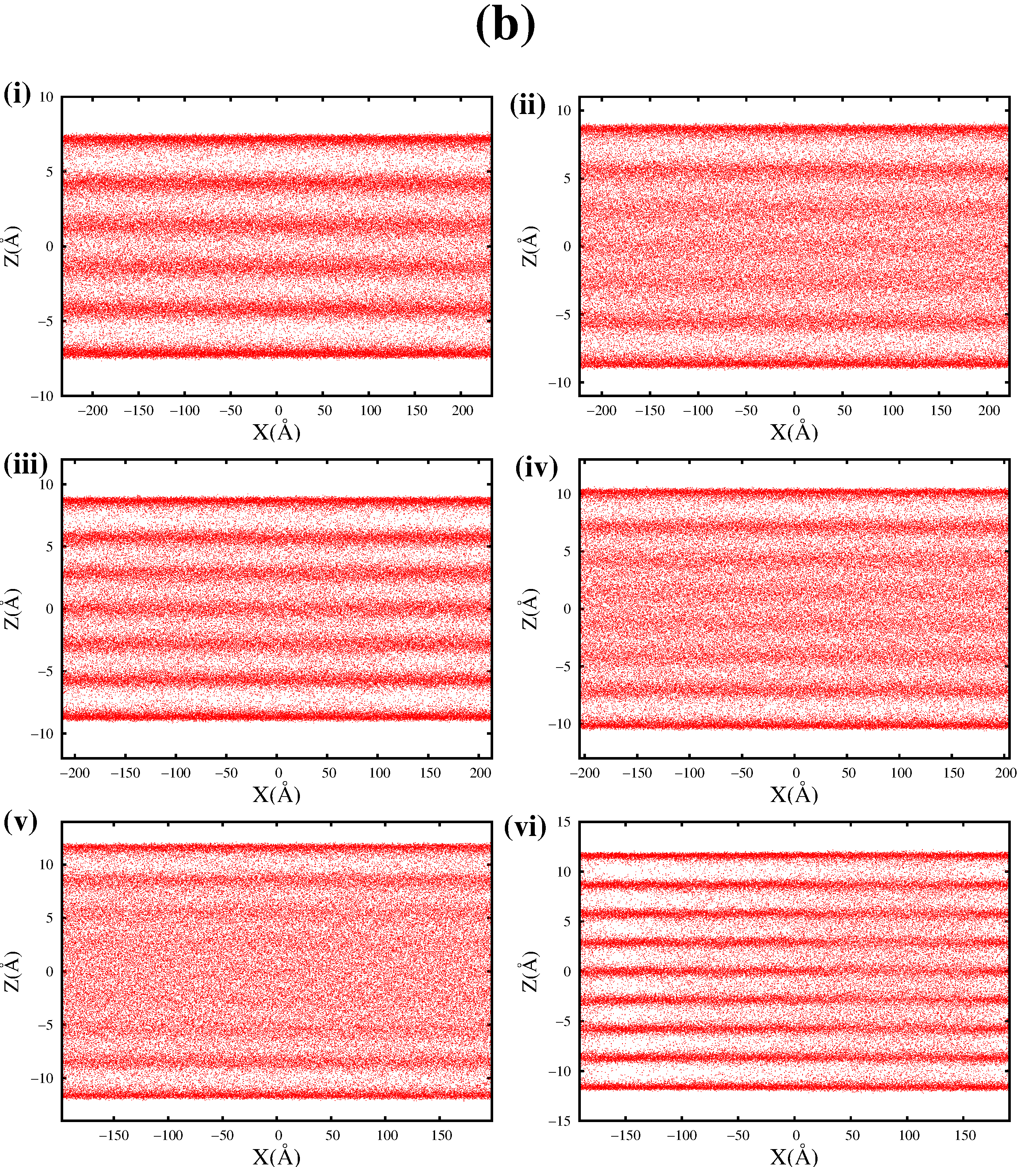}
\end{minipage}
\caption{\label{9} \textbf{(a)} Distribution of Argon particles in supercritical state, averaged over several timesteps at $300$K normal to the walls for a range of confined spacings($20$ $\AA$ $\leqslant$ H $\leqslant$ $30$ $\AA$). \textbf{(b)} Two-dimensional($2$D) projections ($X$-$Z$) of particle configurations for the same set of confined spacings:(i)H=$20$$\AA$, (ii)H=$22$$\AA$, (iii)H=$24$$\AA$, (iv)H=$26$$\AA$, (v)H=$28$$\AA$, (vi)H=$30$$\AA$.}
\end{figure}
\end{center}
\end{widetext}
Fig.\ref{9} shows the distribution profile (Fig.\ref{9}.(a)) and the two-dimensional projections of particle configurations (Fig.\ref{9}.(b)) for a specific range of confined spacings ($20$ $\AA$ $\leqslant$ H $\leqslant$ $30$ $\AA$). Spacings with the ratio of $\frac{H}{\sigma}$ close to integer values (H = $20$ $\AA$, $24$ $\AA$ and $30$ $\AA$) appear to accommodate particles in layers characterized by large amplitude oscillations with the number of peaks scaling linearly with $\frac{H}{\sigma}$. The corresponding two-dimensional projections of particle distribution show well-formed periodic structures. When the spacings are not close to the integer values of $\frac{H}{\sigma}$ (H = $22$ $\AA$, $26$ $\AA$ and $28$ $\AA$), the layers, particularly in the central region of the width, exhibit weak oscillations, due to the frustration involved in accommodating the particles. The corresponding two-dimensional projection of particle configuration clearly shows the breakdown of ordered patterns around the central region (near $z$ = $0$). It is of interest to note that well-developed layers seem to form when $\frac{H}{\sigma}$ is precisely an integer for hard spheres under confinement \cite{Mittal2007} for certain packing fractions. However, in our study we observe that not all integer multiple of $\sigma$ spacings between the walls favour distinct ordering of supercritical fluid particles. While H = $7$$\sigma$ ($23.8$ $\AA$) shows well-developed layering structure, H = $8$$\sigma$ ($27.2$ $\AA$) doesn't show prominent layering (Fig.\ref{10}). After H = $30$ $\AA$, the distribution is nearly featureless in the central region around z = $0$, corresponding to the bulk average density (Fig.\ref{11}). Thus H = $30$ $\AA$ appears to indicate a sharp transition in particle number distributions along z. 

\subsubsection{\label{sec:level3}\textbf{Translational order parameter and Two-body excess entropy studies : Quantification of ordering normal to the walls}}
Although the particle distribution normal to the walls before crossing the Frenkel line clearly shows the ordering of the particles perpendicular to the walls due to confinement, there are several other measures that help to quantify the structure. Two such measures are the translational order parameter ($\tau$) and two-body excess entropy ($s^{(2)}$), both of which can be determined by the radial distribution function normal to the walls denoted by $g_{\perp}(r)$. $g_{\perp}(r)$ can be defined as 
\begin{align}
g_{\perp}(r) \equiv & \frac{1}{\rho^2V} \sum_{i\neq j}\delta \left(r -r_{ij}\right)\Bigg[\theta\left(|x_{i}-x_{j}|+\frac{\delta x}{2}\right)\nonumber\\
& -\theta\left(|x_{i}-x_{j}|-\frac{\delta x}{2}\right)\Bigg]
\end{align}

where $V$ is the volume, $\rho$ is mass density, $r_{ij}$ is the distance normal to the walls between molecules i and j, $x_i$ is the $x$ coordinate of the molecule $i$, and $\delta(x)$ is the Dirac $\delta$ function. The Heaviside function $\theta(x)$ restricts the sum to a pair of particles located in the same slab of thickness $\delta x$. Translational order parameter ($\tau$), which can be used as a tool to probe local density modulations in a system, is defined as \cite{Pant2016, Truskett2000}
\begin{equation}
\tau = \int_0^{r_{c}} |g_{\perp}(r)-1|dr
\end{equation}
where, $r_c$ is the numerical cut-off of the RDF, along normal plane to the wall. For a completely uncorrelated systems like ideal gas $\tau$ is zero as $g_{r}$ $\equiv$ $1$ for such systems. As system becomes ordered and structured, its value becomes relatively large \cite{Pant2016, Truskett2000}. We calculate $\tau$ from the $g_{\perp}(r)$ for both the confined and bulk systems before and after crossing the Frenkel line. Fig.\ref{12}.(a) shows the translational order parameter ($\tau$) variation for different confined spacings along the Frenkel line of supercritical Argon. For comparison, the corresponding values for bulk Argon in supercritical phase have been computed.\\

Another estimate of the structure of the confined system is the excess entropy, defined as the difference between the entropy of the probed system and the ideal gas calculated at a same density, temperature combination. We use the two-body excess entropy $s^{(2)}$, defined \cite{Nettleton1958, Mountain1971} as
\begin{equation}
s^{(2)} = -2 \pi \rho k_B \int \left[g_{\perp}(r)lng_{\perp}(r)-g_{\perp}(r)+1\right]r^2dr
\end{equation}
As $g_{\perp}(r)$ $\to$ $1$ for completely uncorrelated and disordered systems, $s^{(2)}$ vanishes(ideal gas behaviour) and becomes large and negative as an ordered structure starts forming \cite{Truskett2000}. Fig.\ref{12}.(b) shows the two-body excess entropy as a function of confined spacings for the supercritical Argon along the Frenkel line. \\
For small spacings, the values of $\tau$ and the scaled $s^{(2)}$ can be seen to be respectively higher and lower than the corresponding bulk phase values. It is worth noting that $\tau$ for bulk phases before and after the Frenkel line are positive, indicating that the supercritical bulk phases are more ordered than the ideal gas phase. Similarly, the bulk values of scaled $s^{(2)}$ show non-zero negative values , which indicates that some ordering is present even in bulk with respect to the ideal gas structure. The variation of the scaled $s^{(2)}$ with confinement is arising purely from ordering since the density has been kept constant for all confinements at a chosen P,T point.
\begin{widetext}
\begin{center}
\begin{figure}[H]
\begin{minipage}{.5\textwidth}
\includegraphics[width=1.0\textwidth]{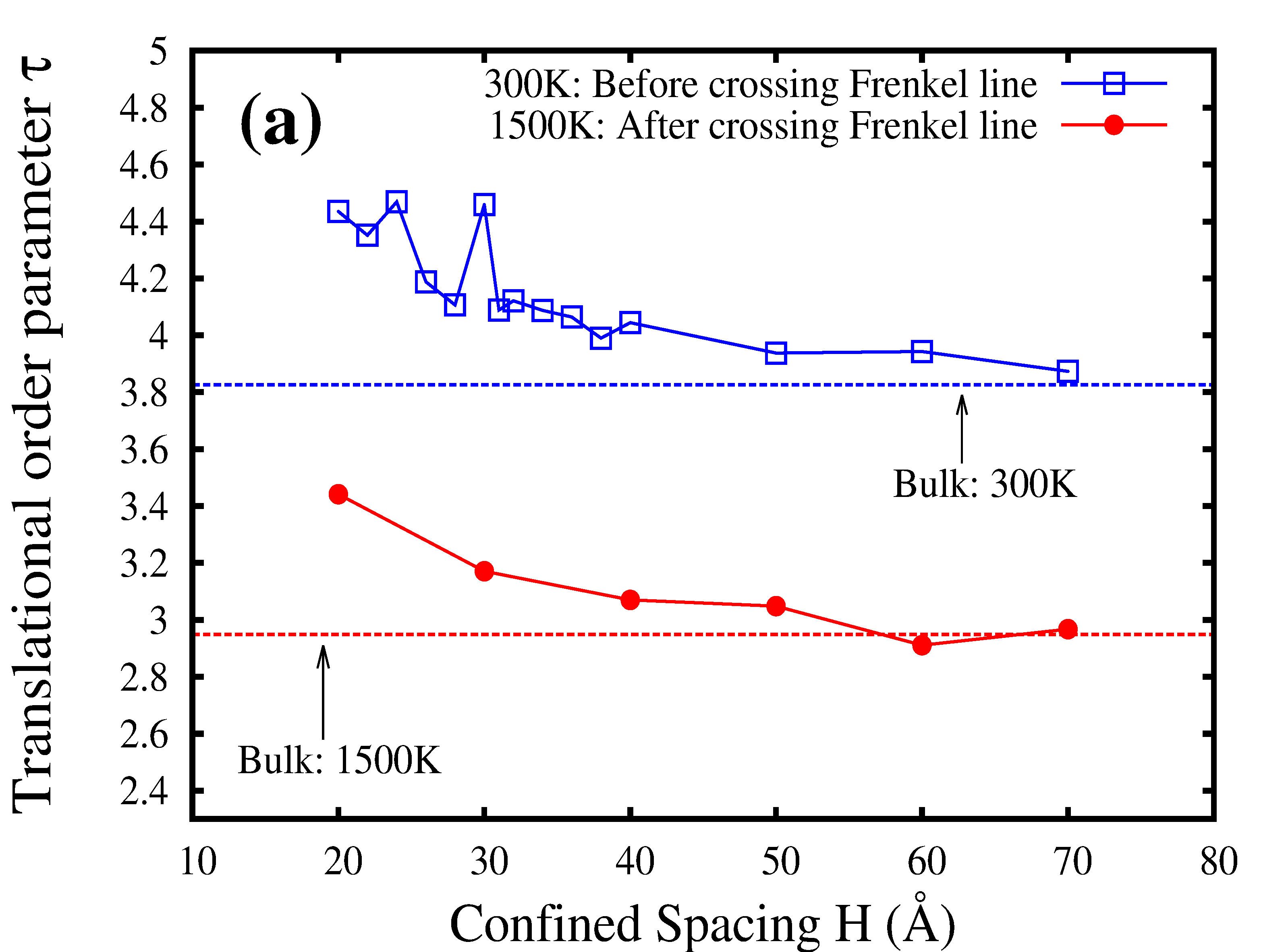}
\end{minipage}
\begin{minipage}{.5\textwidth}
\includegraphics[width=1.0\textwidth]{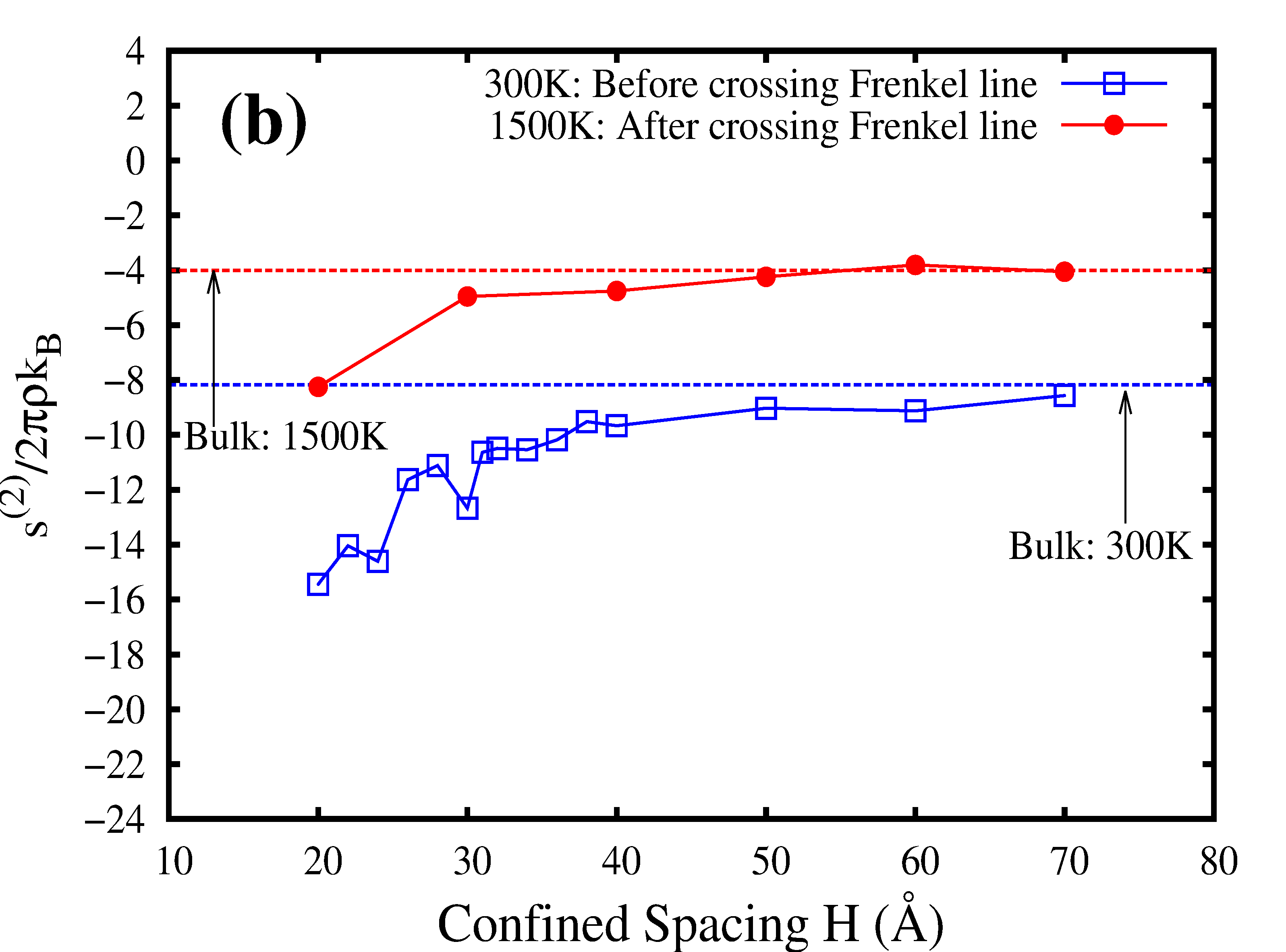}
\end{minipage}
\caption{\label{12} \textbf{(a)} Translational order parameter ($\tau$) variation with different confined spacings both before and after crossing the Frenkel line of supercritical Argon. For reference the bulk phase values have been shown. \textbf{(b)} Scaled two-body excess entropy ($s^{(2)}/2 \pi \rho k_B$) variation with different confined spacings both before and after Frenkel line of supercritical Argon. The corresponding Bulk phase values have also been shown for reference.}
\end{figure}
\end{center}
\end{widetext}

\begin{figure}[H]
\centering
\includegraphics[width=0.5\textwidth]{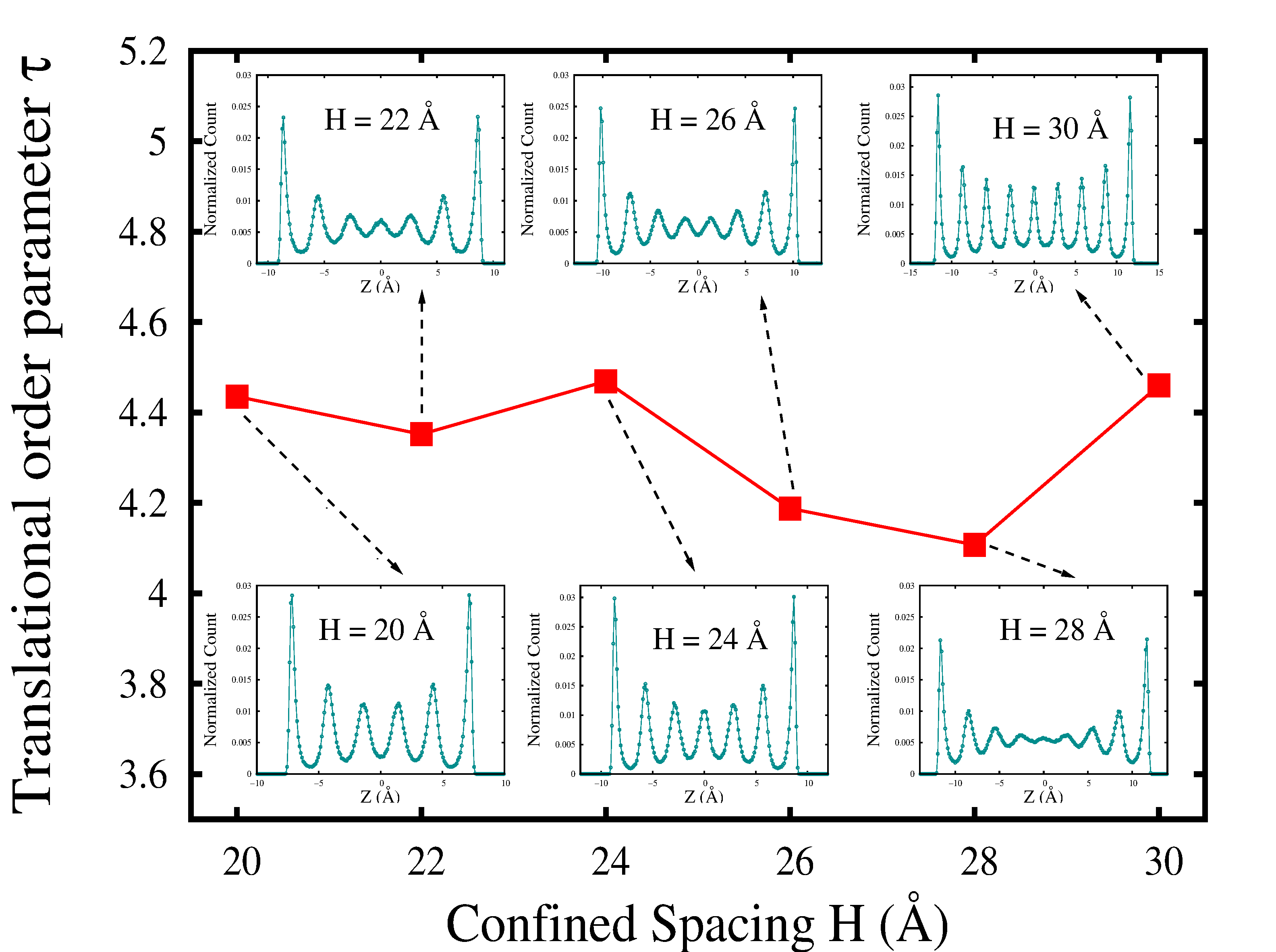}
\caption{\label{13} Pictorial representation of the in-phase behaviour between the variation of $\tau$ and sharply developed layers along $z$, under confinement, for state point before crossing Frenkel line ($300$K). Higher values of $\tau$ correspond to the sharply developed layers.}
\end{figure}
In Fig.\ref{12}.\textbf{(a)} and Fig.\ref{12}.\textbf{(b)} we show the variation of $\tau$ and $s^{(2)}$ with different confined spacings. We observe oscillating behaviour of both $\tau$ and $s^{(2)}$ under the atomistic boundary confinement for a range of spacings (H $\leqslant$ $30$ $\AA$) at $300$K. Spacings leading to well developed sharper layers (comparatively lesser packing frustration) gives rise to higher values of $\tau$ and lower values of $s^{(2)}$, in contrast to the spacings where under-developed layers are formed (packing frustration dominates). Fig.\ref{13} pictorially depicts these features. 

The spacing (H = $30$ $\AA$) at $300$K, mentioned earlier, shows distinct jumps in both $\tau$ and $s^{(2)}$ indicating significant loss of ordering. For spacings beyond H = $30$ $\AA$, $\tau$ and $s^{(2)}$ can be seen to tend to their respective bulk values. In phase "B"(after crossing the Frenkel line) $\tau$ and $s^{(2)}$ indicate mild ordering only for very small spacings. 

\subsubsection{\label{sec:level3}\textbf{Structural features parallel to the walls : Before and after crossing the Frenkel line}}
At $300$K we have seen that confinement leads to layering over the width of the spacing. To understand the effect of the confinement on the structural behaviour of supercritical fluid parallel to the walls, we examine the parallel components of RDF for each of these layers. The radial distribution function for each of these layers can be evaluated from \cite{Pradeep2005, Krott2013}
\begin{align}
g_{\parallel}(r) \equiv & \frac{1}{\rho^2V} \sum_{i\neq j}\delta \left(r -r_{ij}\right)\Bigg[\theta\left(|z_{i}-z_{j}|+\frac{\delta z}{2}\right)\nonumber\\
& -\theta\left(|z_{i}-z_{j}|-\frac{\delta z}{2}\right)\Bigg]
\end{align}
where $V$ is the volume, $\rho$ is the mass density, $r_{ij}$ is the distance parallel to the walls between molecules i and j, $z_i$ is the $z$ coordinate of the molecule $i$, and $\delta(x)$ is the Dirac $\delta$ function. The Heaviside function $\theta(x)$ restricts the sum to a pair of particles located in the same slab of thickness $\delta z$. We have considered $\delta z$ to be same as the width of each layer. We use a uniform bin width and bin number of $80$ to calculate $g_{\parallel}(r)$ for all the cases. Two well-defined classes of $g_{\parallel}(r)$, henceforth labelled "P" and "Q", can be seen from Fig.\ref{14}.(a) and (b) respectively. $g_{\parallel}(r)$ for Class "P" is characterized by the absence of positional shift of the peak positions for different layers. 
\begin{widetext}
\begin{center}
\begin{figure}[H]
\begin{minipage}{0.5\textwidth}
\includegraphics[width=0.87\textwidth]{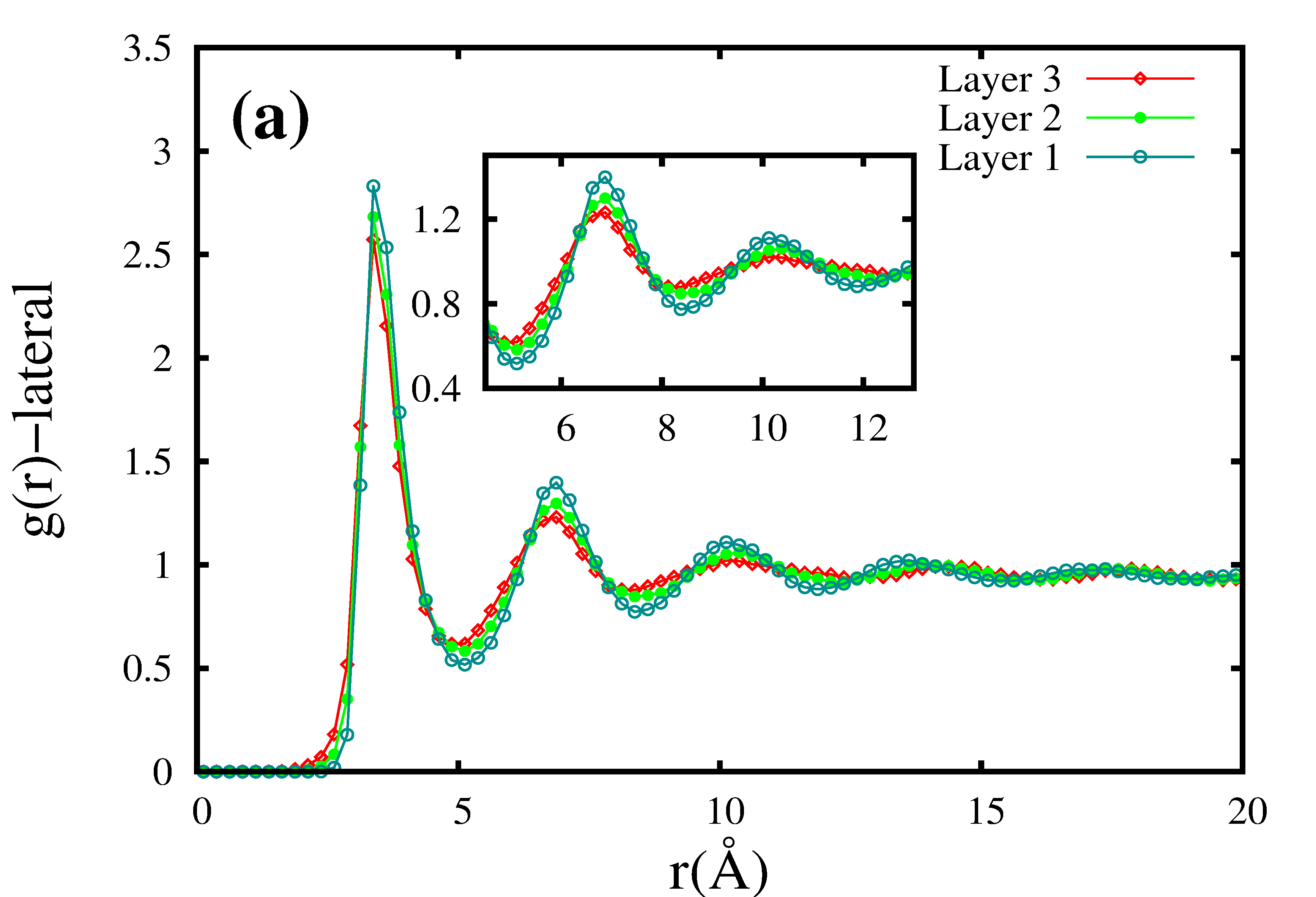}
\end{minipage}
\begin{minipage}{0.5\textwidth}
\includegraphics[width=0.87\textwidth]{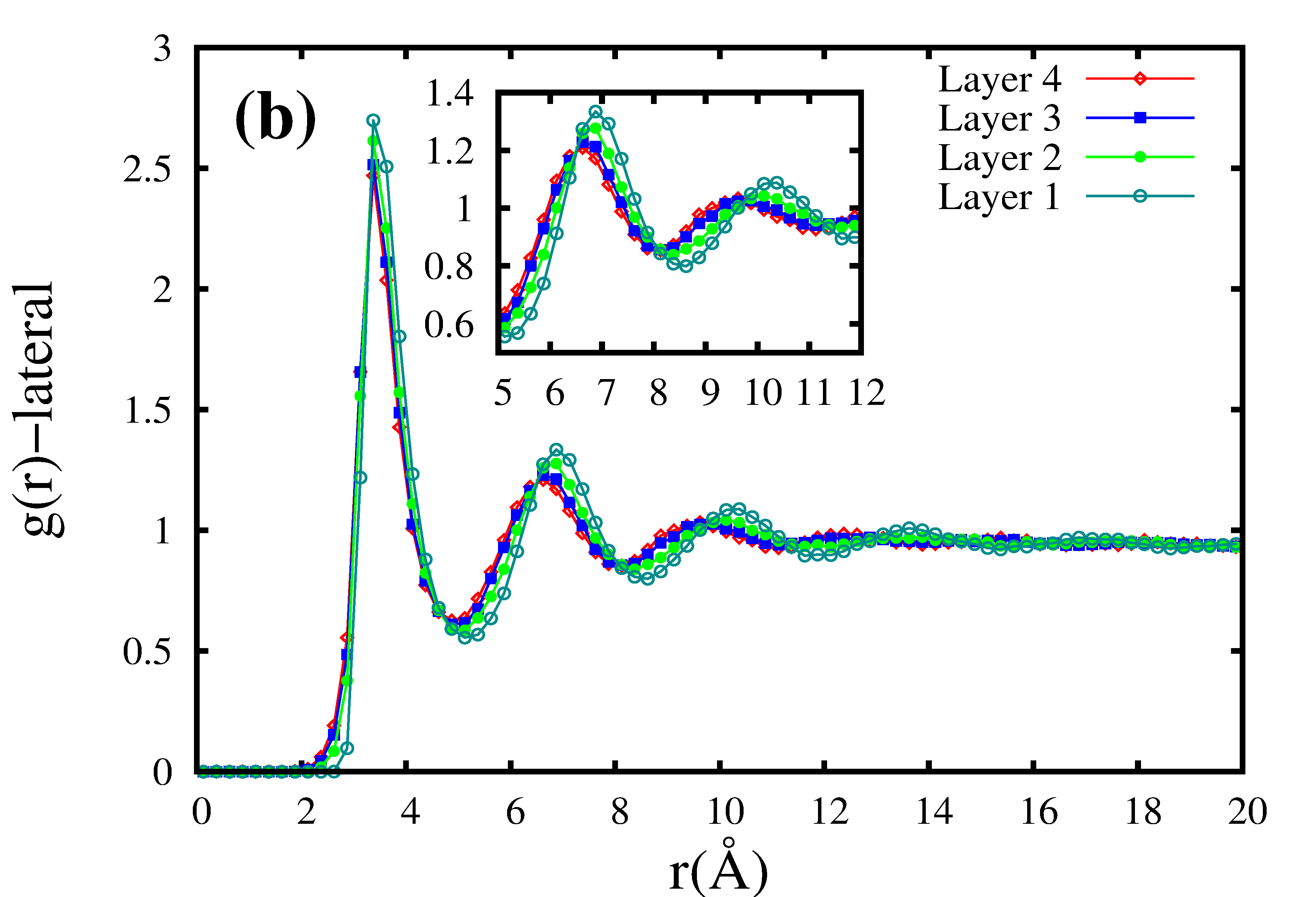}
\end{minipage}
\caption{\label{14} $g_{\parallel}(r)$ of Argon particles in supercritical phase under different confined spacings (H) at $300$K parallel to the walls. \textbf{(a)}H=$20$ $\AA$ (class P), \textbf{(b)}H=$26$ $\AA$ (class Q). Insets provide the details of the second and third RDF peaks parallel to the walls. By symmetry, only one set of layers with respect to the centre is considered. Numbering of layers starts from the layer closest to the wall.}
\end{figure}
\end{center}
\end{widetext}

\begin{figure}[H]
\centering
\includegraphics[width=0.5\textwidth]{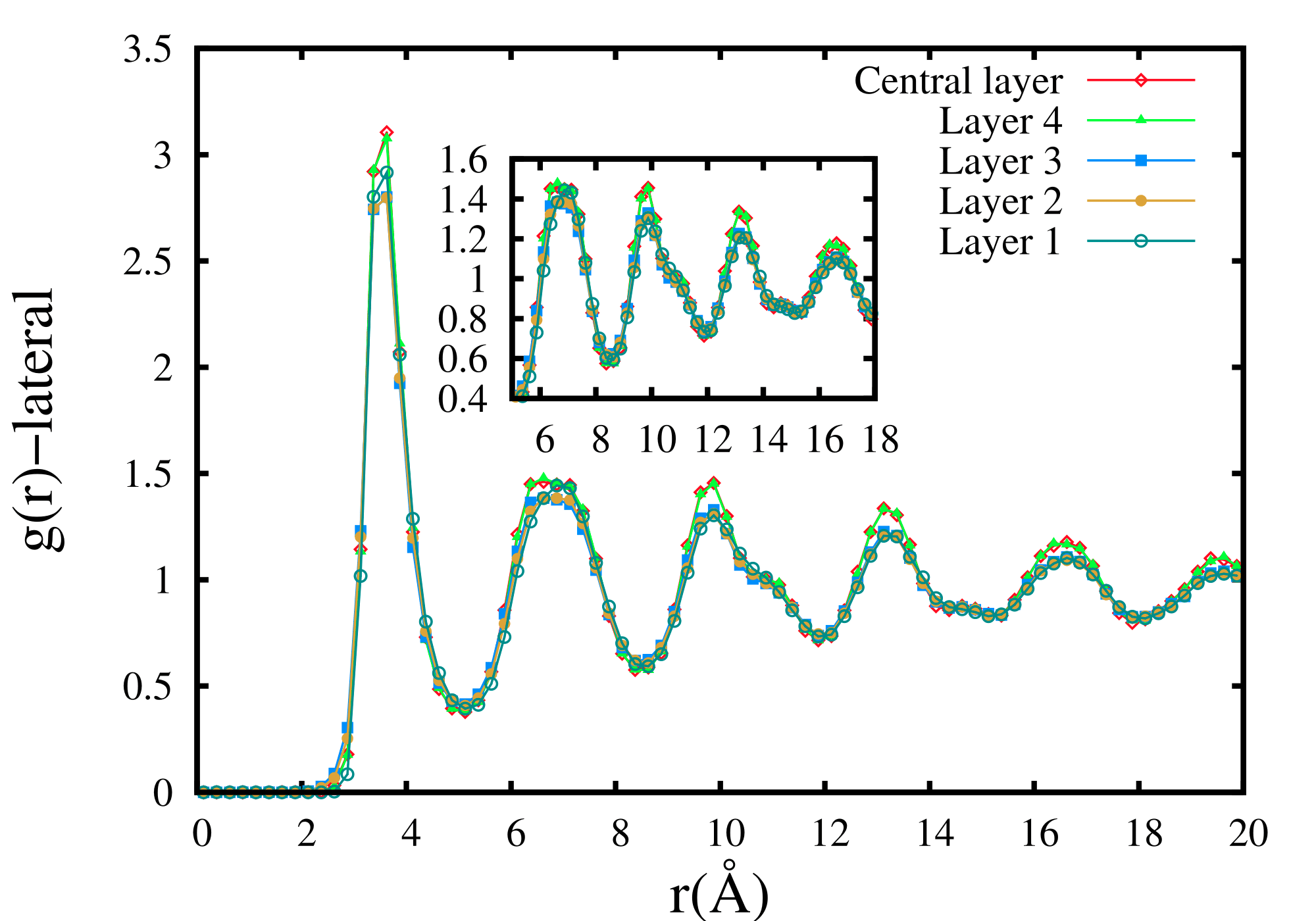}
\caption{\label{15} $g_{\parallel}(r)$ of Argon particles in supercritical phase at $300$K for H=$30$ $\AA$ spacing. Only one set of layers with respect to the centre is considered. Numbering of layers starts from the layer closest to the wall. Inset shows the absence of lateral shift of RDF peaks (class P).}
\end{figure}
However, $g_{\parallel}(r)$ for Class "Q" shows a gentle shift in the peak positions for different layers. With respect to the central layer the positional shift gradually increases and reaches a maximum for the layer closest to the wall. We note that class "Q" deals with confinement spacings which are not nearly equal to the integer multiple of $\frac{H}{\sigma}$ leading to packing frustration. It has been found that $g_{\parallel}(r)$ for H $\geqslant$ $30$ $\AA$ belongs to class "Q".\\
Confinement with H = $30$ $\AA$ shows an additional feature although it belongs to class "P": it shows amorphous-like structure formation parallel to the walls (Fig.\ref{15}). Well-defined coordination spheres upto a length of $20$ $\AA$ ($\approx$ $6$ $\sigma$) parallel to the walls, observed in $g_{\parallel}(r)$, clearly shows more ordering than a typical "liquidlike"
\begin{figure}[H]
\includegraphics[width=0.5\textwidth]{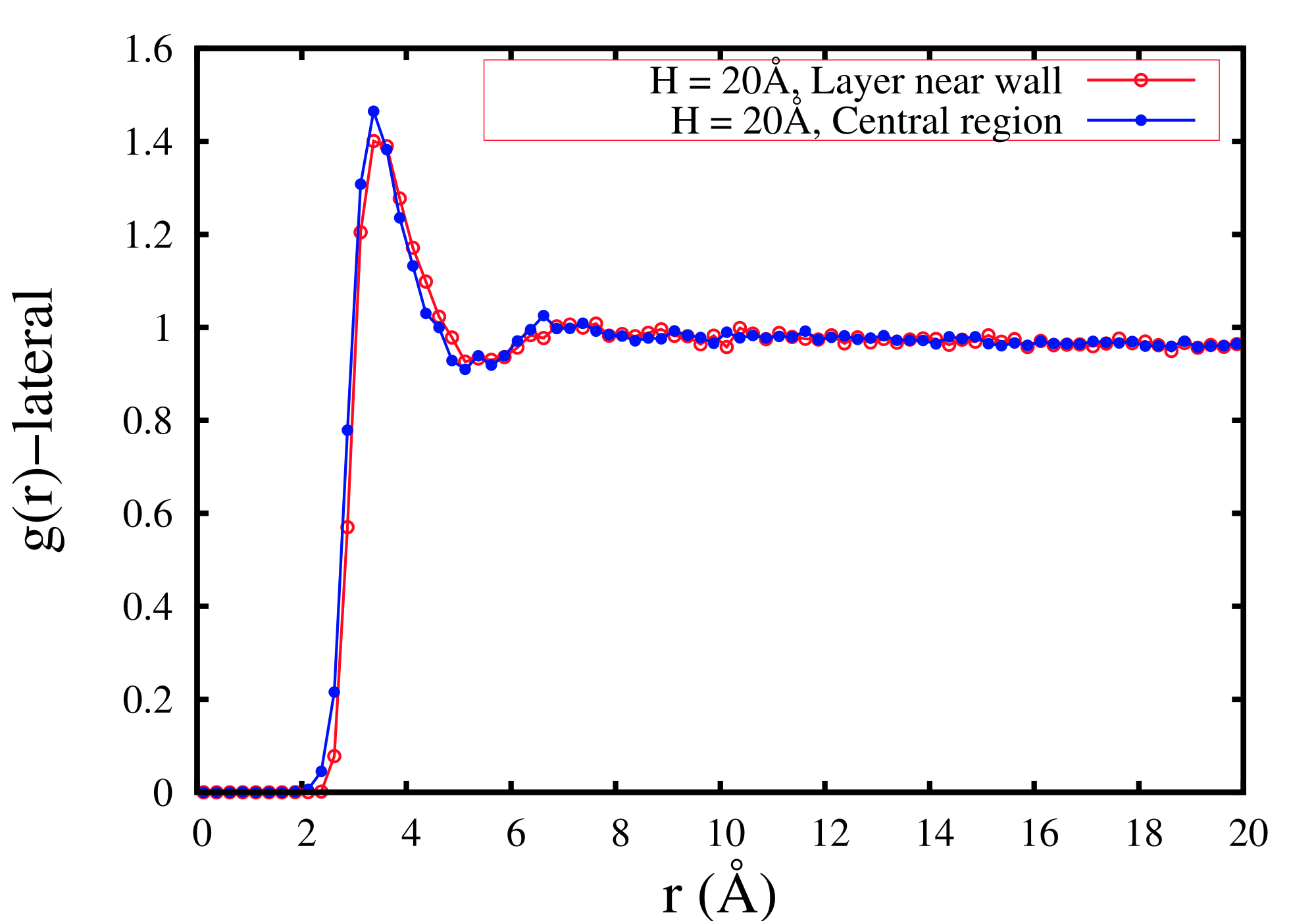}
\caption{\label{16} The nature of lateral RDF ($g_{\parallel}(r)$)of central region and region close to the walls after crossing Frenkel line at $1500$K temperature for H = $20$ $\AA$.}
\end{figure}
phase, where local ordering persists only upto a maximum of $2$ to $3$ coordination spheres.\\
At $1500$K (after crossing the Frenkel line), we don't see distinct structural features parallel to the walls. $g_{\parallel}(r)$ exhibits only a single peak, shown in Fig.\ref{16} reminiscent of gaslike phase. It may be noted that $g_{\parallel}(r)$ for the layer closest to the wall is no different from the $g_{\parallel}(r)$ of the central region.
\subsubsection{\label{sec:level3}\textbf{Structural features of supercritical fluid for sufficiently narrow spacings under confinement: Before and After crossing the Frenkel line }}
\begin{figure}[H]
\begin{minipage}{.5\textwidth}
\includegraphics[width=0.74\textwidth]{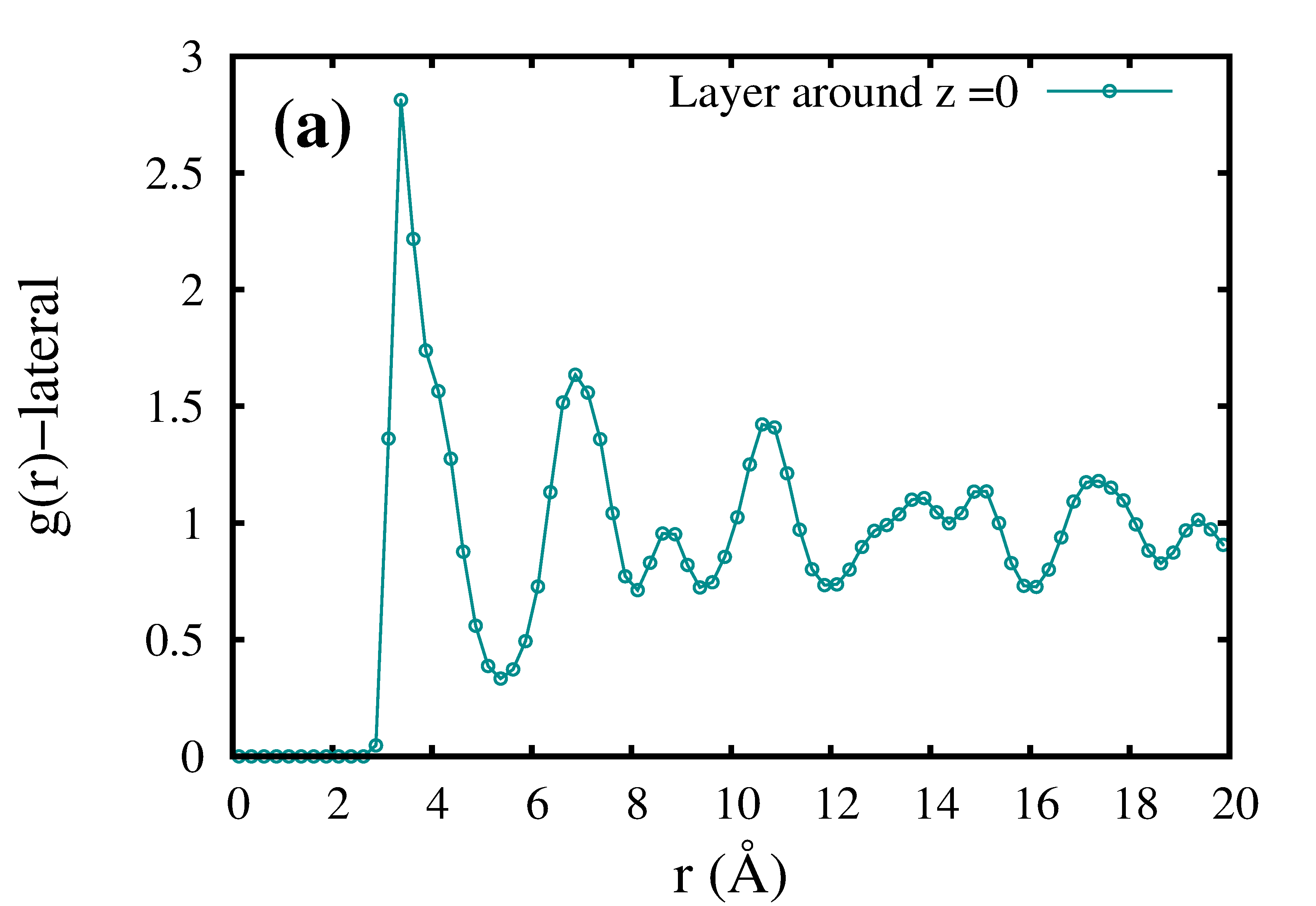}
\end{minipage}
\begin{minipage}{.5\textwidth}
\includegraphics[width=0.74\textwidth]{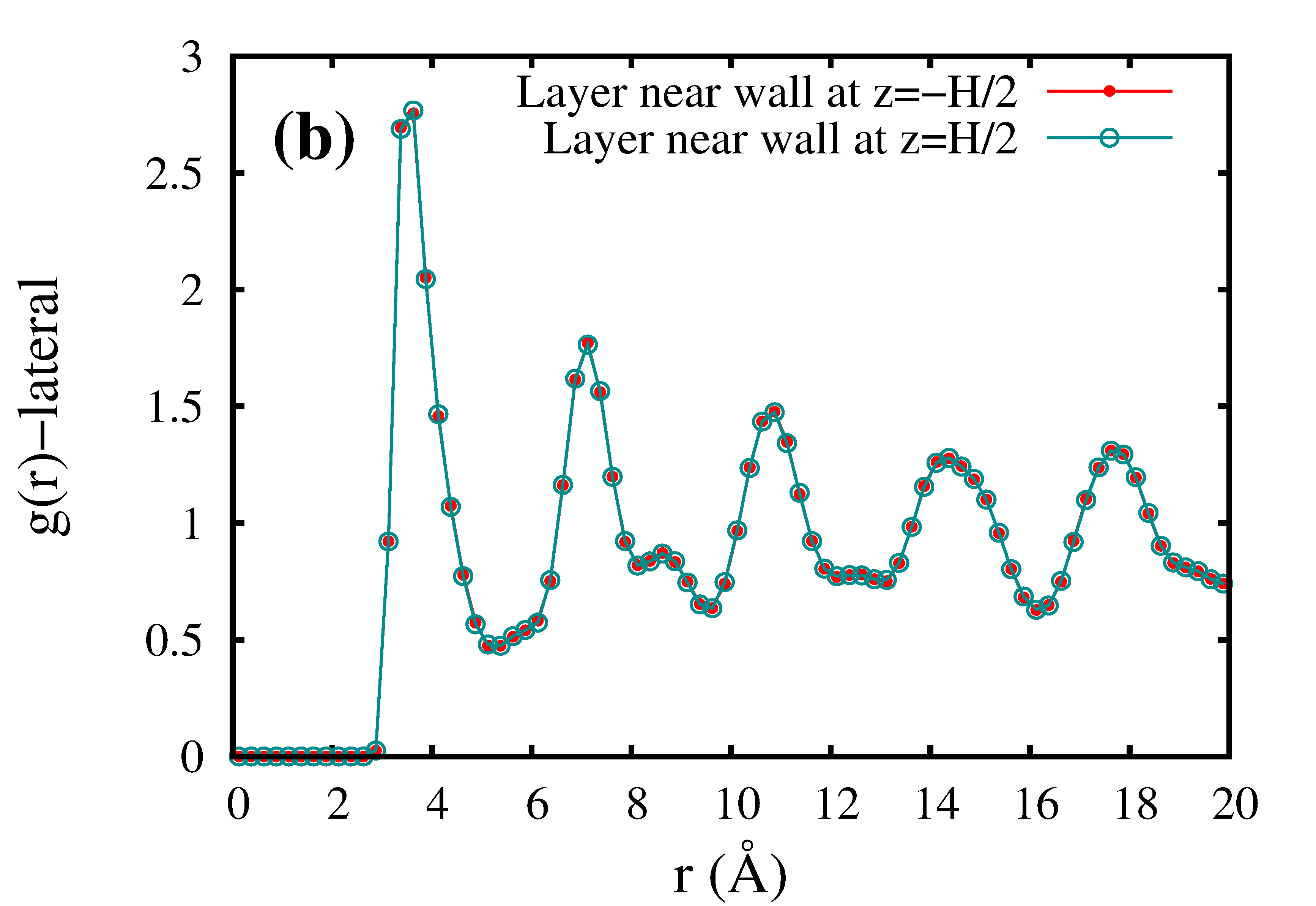}
\end{minipage}
\begin{minipage}{.5\textwidth}
\includegraphics[width=0.74\textwidth]{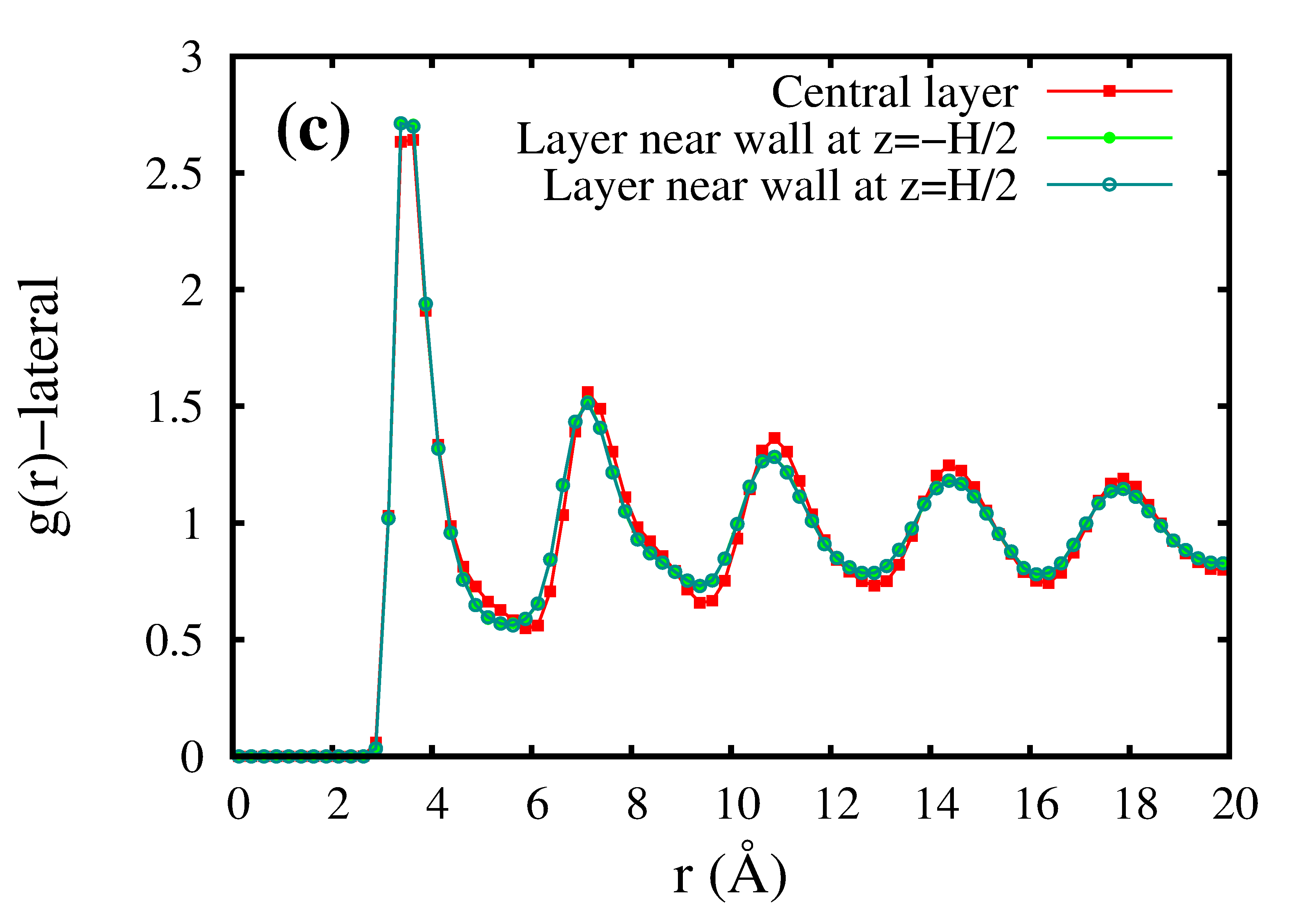}
\end{minipage}
\caption{\label{18} Radial Distribution function ($g_{\parallel} (r)$) of Argon particles in supercritical phase under different narrow confined spacings (H) at $300$K parallel to the walls.\textbf{(a)}H=$3.4$ $\AA$, \textbf{(b)}H=$6.8$ $\AA$, \textbf{(c)}H=$10$ $\AA$.}
\end{figure}

\begin{figure}[H]
\centering
\includegraphics[width=0.45\textwidth]{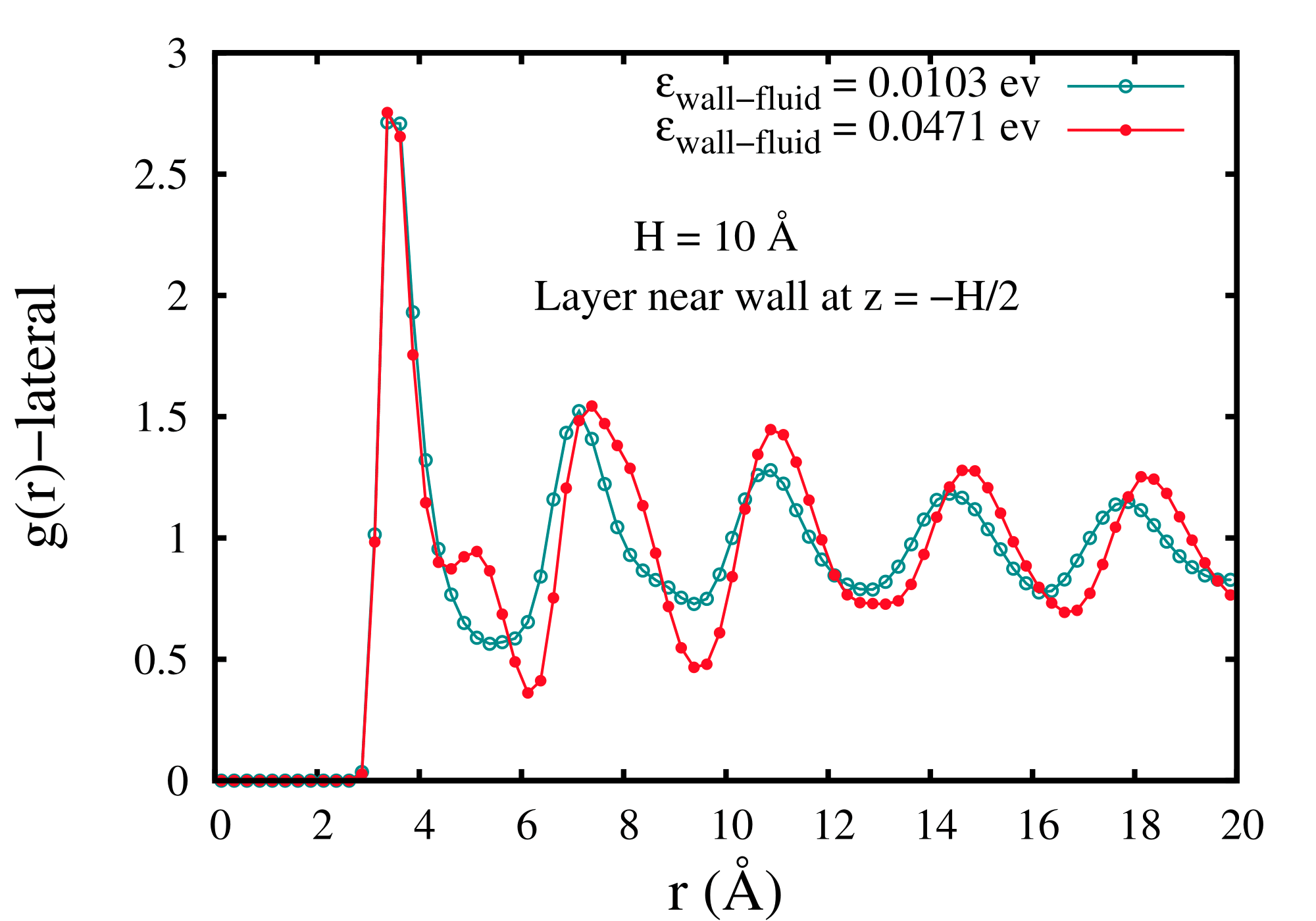}
\caption{\label{19} Lateral Radial Distribution function ($g_{\parallel}(r)$) of supercritical Argon for H = $10$ $\AA$ at $300$K parallel to the walls. Two different LJ interaction parameters($\epsilon_{wall-fluid}$) $0.0103$ ev and $0.0471$ ev have been used for comparison. $g_{\parallel}(r)$ for layers close to the wall at z =-$H/2$ have been chosen for comparison.}
\end{figure}

We investigate the structural features of supercritical Argon for extremely tight confinements ($3.4$ $\AA$ $\leqslant$ H $\leqslant$ $10$ $\AA$). 
\begin{figure}[H]
\begin{minipage}{.5\textwidth}
\includegraphics[width=0.74\textwidth]{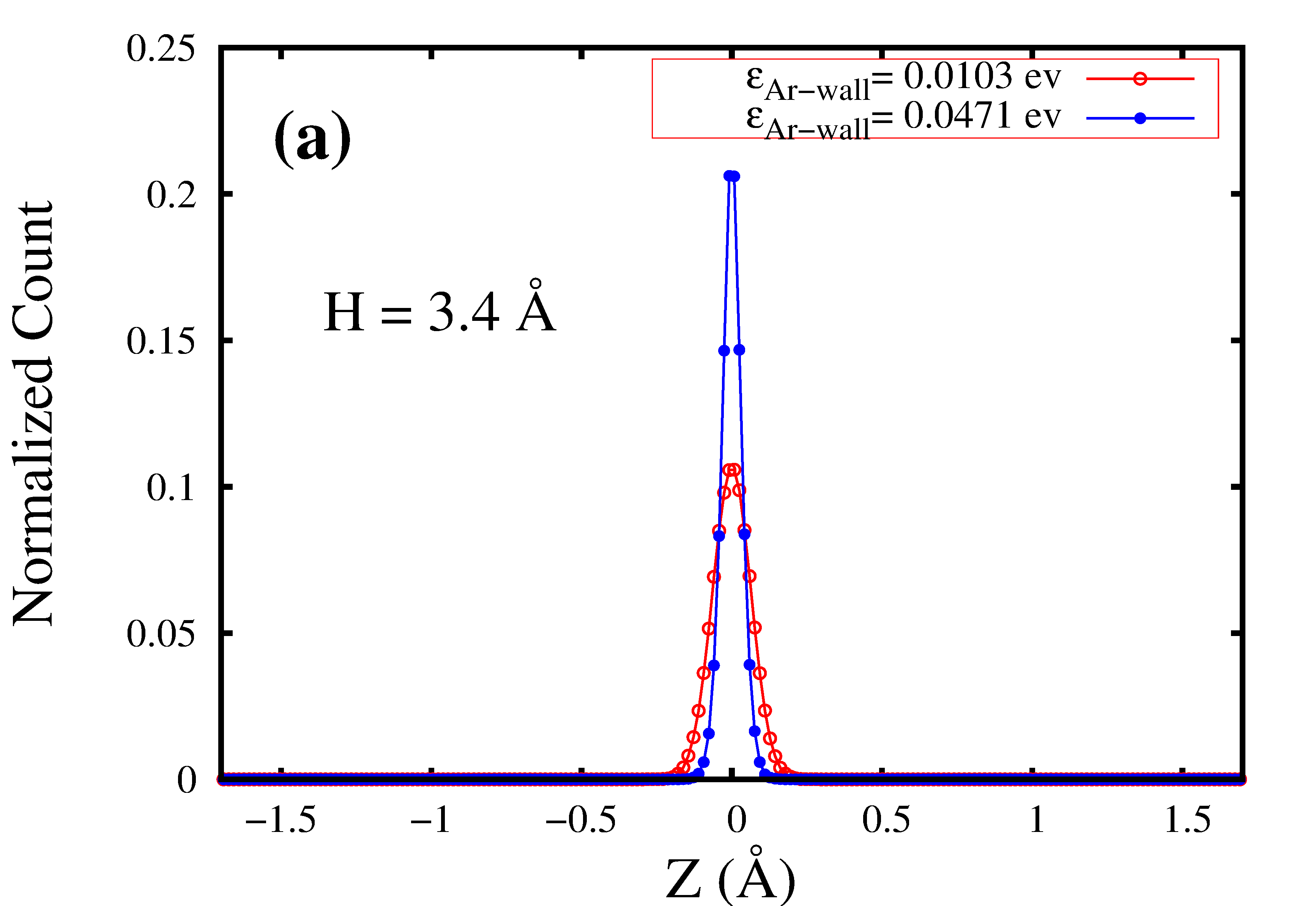}
\end{minipage}
\begin{minipage}{.5\textwidth}
\includegraphics[width=0.74\textwidth]{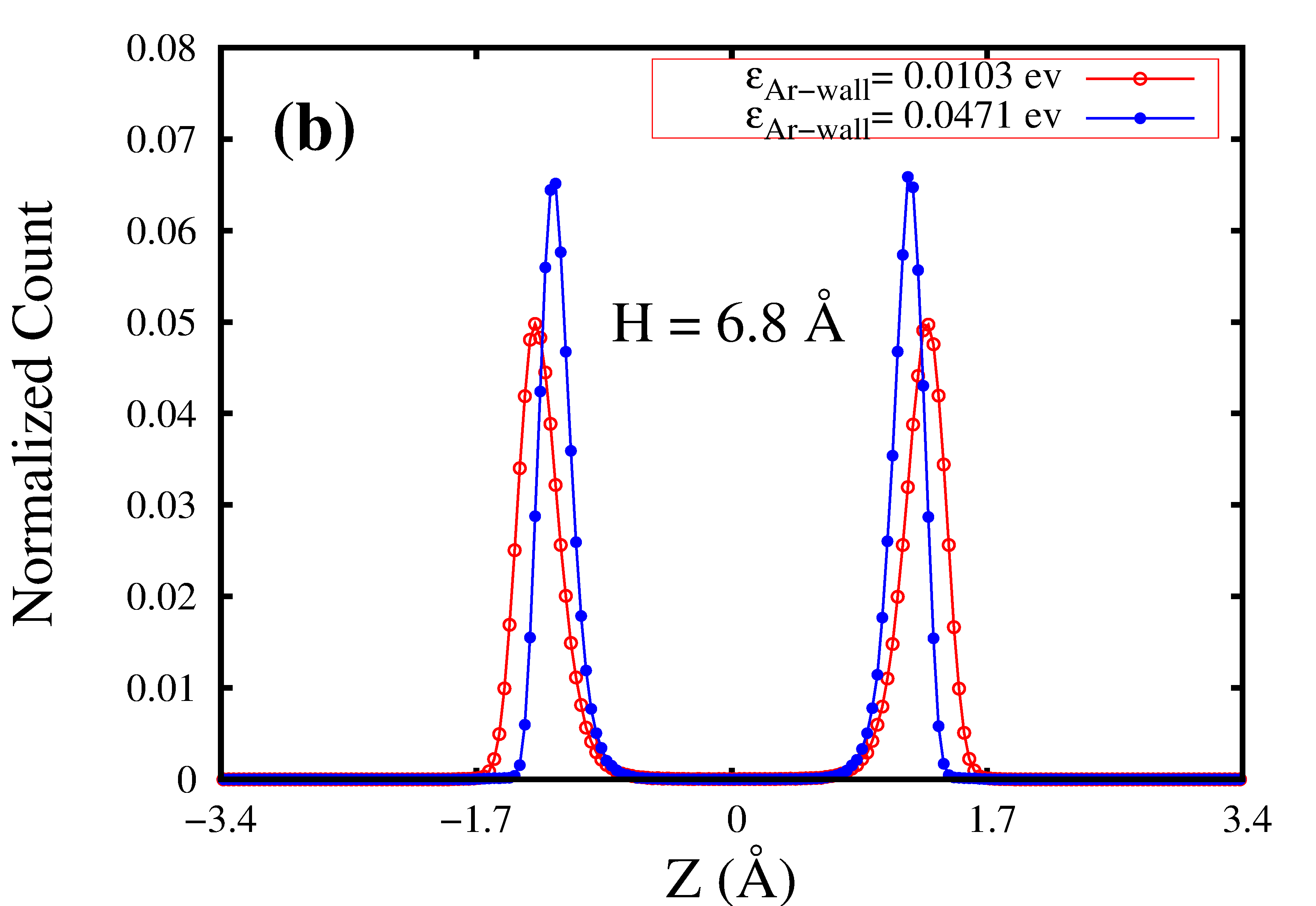}
\end{minipage}
\begin{minipage}{.5\textwidth}
\includegraphics[width=0.74\textwidth]{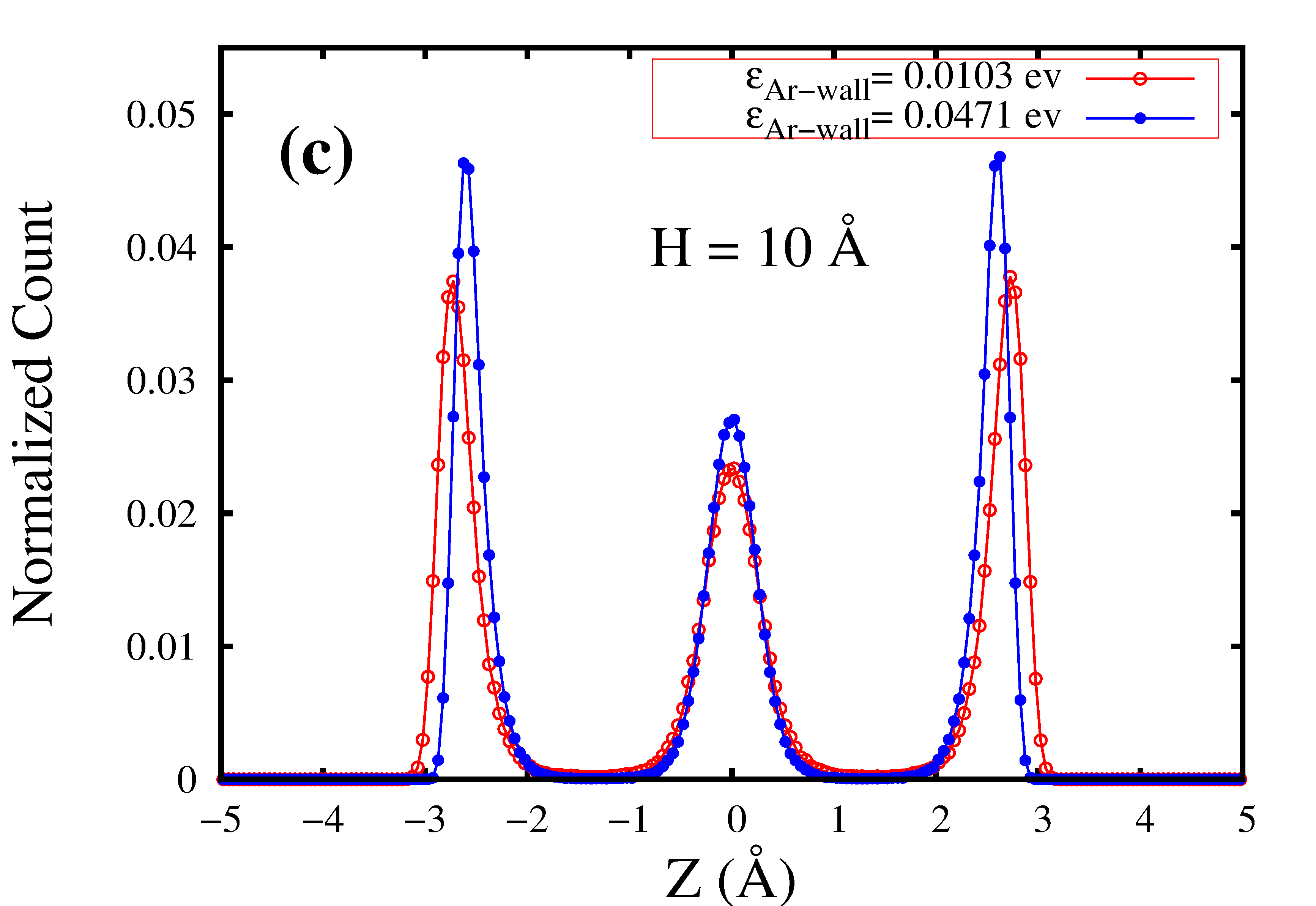}
\end{minipage}
\caption{\label{17} Distribution of Argon particles in supercritical state, averaged over several timesteps at $300$K normal to the walls for sufficiently narrow confined spacings(H).\textbf{(a)} H=$3.4$ $\AA$, \textbf{(b)}H=$6.8$ $\AA$, \textbf{(c)}H=$10$ $\AA$. Two different LJ interaction parameters($\epsilon_{wall-fluid}$) $0.0103$ ev and $0.0471$ ev have been used.}
\end{figure}
Fig.\ref{17} shows the distinct layering of the particles in supercritical state at $300$K, before crossing the Frenkel line. The number of layers formed due to confinement, scales linearly with the ratio of spacing to the atomic diameter. For example, a spacing of $1$ atomic diameter(H=$3.4$ $\AA$) gives $1$ peak, $2$ atomic diameters(H=$6.8$ $\AA$) give $2$ peaks and so on. Successive peaks are spaced lower than the atomic diameter of a particle, on average. Thus, despite having prominent layer arrangement of particles as seen from the number distribution profile, it is different from the close-packed structure.\\
To examine the role of the wall-fluid interaction potential, simulations are carried out with higher interaction strength($\epsilon$=$0.0471$ev). Fig.\ref{17} shows that, while, the general features of layering remain unaltered, there is a shift in the peak location and an enhancement of the peak heights.\\
Fig.\ref{18} shows $g_{\parallel}(r)$ for each of these layers, exhibits
 structural features corresponding to amorphous phase at $300$K. This property has also been seen in the case of $g_{\parallel}(r)$ for H = $30$ $\AA$. Fig.\ref{18}.(b) and (c) show almost identical trends in $g_{\parallel}(r)$ for different layers, corresponding to spacings of $6.8$ $\AA$ and $10$ $\AA$ respectively. \\
Increasing the wall-fluid interaction strength leads to enhanced peak-heights in $g_{\parallel}(r)$ as can be seen from Fig.\ref{19}.In Fig.\ref{20}.(a) a prominent central peak is seen for $3.4$ $\AA$ spacing at $1500$K, after crossing the Frenkel line. For all other spacings at $1500$K, we observe weak layering near the wall and almost flat distribution in between. $g_{\parallel}(r)$ for all spacings, including H = $3.4$ $\AA$, show structural features resembling the gas phase (Fig.\ref{20}.(b)).\\

\subsubsection{\label{sec:level3}\textbf{Evolution of structural features along the isobaric line under strong confinement}}

Until now we have discussed about the structural features under confinement for two state points of supercritical Argon at a pressure of $5000$ bar: One(point A) before crossing the Frenkel line($300$K) and the other(point B) after crossing the Frenkel line($1500$K). Though these two state points A and B, clearly showed distinct liquidlike and gaslike features respectively, it will be interesting to see how the structural features evolve in the vicinity of the Frenkel line, along the isobaric path at $5000$bar. We investigate the structural features of supercritical Argon for various state points close to the Frenkel line under the confinement with atomistic boundaries. We take the case of one narrow confined spacing(H = $10$ $\AA$) and closely monitor the transformation of the features of number distribution profiles along $z$ (Fig.\ref{21}.(a)). 

\begin{widetext}
\begin{center}
\begin{figure}[H]
\includegraphics[width=1.0\textwidth]{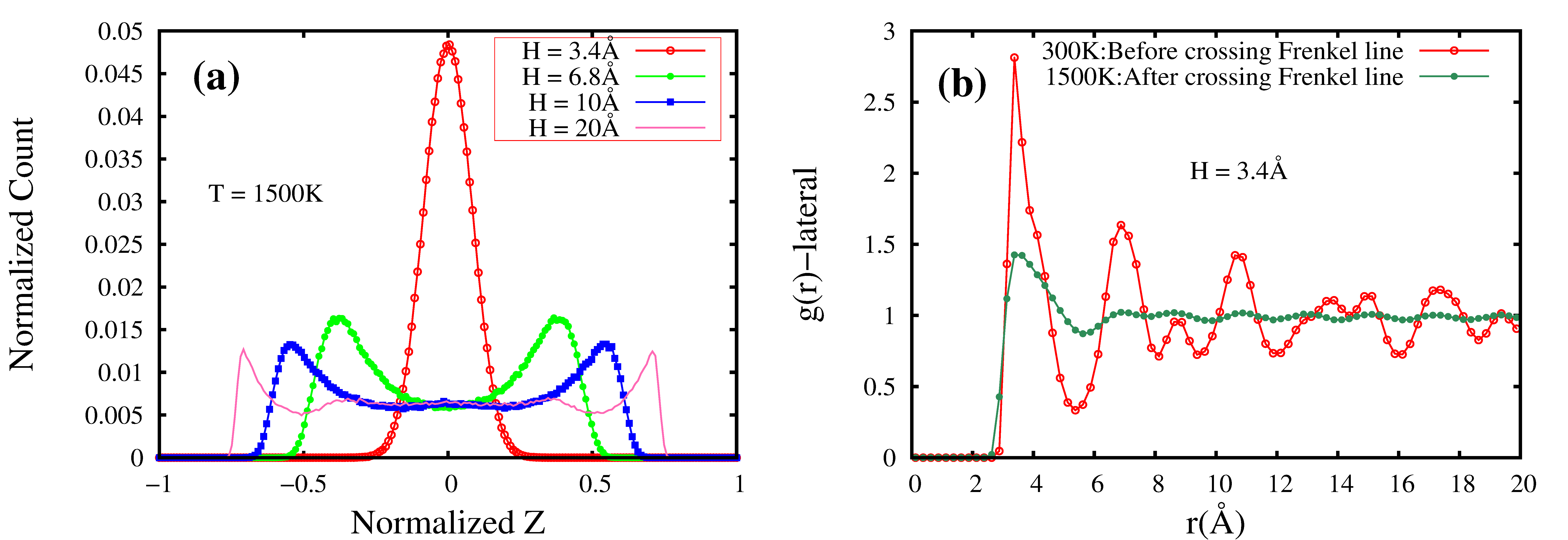}
\caption{\label{20} \textbf{(a)} Distribution of Argon particles in supercritical state, averaged over several timesteps at $1500$K normal to the walls for sufficiently narrow confined spacings. \textbf{(b)} Comparison between ($g_{\parallel}(r)$) of supercritical Argon before($300$K) and after crossing($1500$K) the Frenkel line parallel to the walls for H=$3.4$ $\AA$ spacing.}
\end{figure}
\end{center}
\end{widetext}
In the liquidlike phase at $300$K, three prominent layers are observed, resembling a highly structured fluid. This structural feature gradually decays as we go towards the gaslike phase crossing the Frenkel line. We observe a systematic decay of the peak-height of the central layer from a well defined value to complete disappearance as we go from $300$K to $1500$K along the Frenkel line (Fig.\ref{21}.(a)). This gradual disappearance of the central peak along the Frenkel line($\sim$ $600$K-$700$K) clearly reconfirms the two-phase heterogeneity of supercritical fluids in confinement. The layers close to the walls, though decreased in heights, seem to exist even after crossing Frenkel line due to strong wall-particle correlation near the walls. $g_{\parallel}(r)$ of the central layer also shows gradual transition with multiple coordination spheres(amorphous-like) transform to fewer ($2$,$3$) coordination spheres (liquidlike) and ultimately reduces to a single coordination sphere (gaslike) across the Frenkel line (Fig.\ref{21}.(b))
\onecolumngrid
\begin{widetext}
\begin{center}
\begin{figure}[H]
\begin{minipage}{0.5\textwidth}
\includegraphics[width=1.0\textwidth]{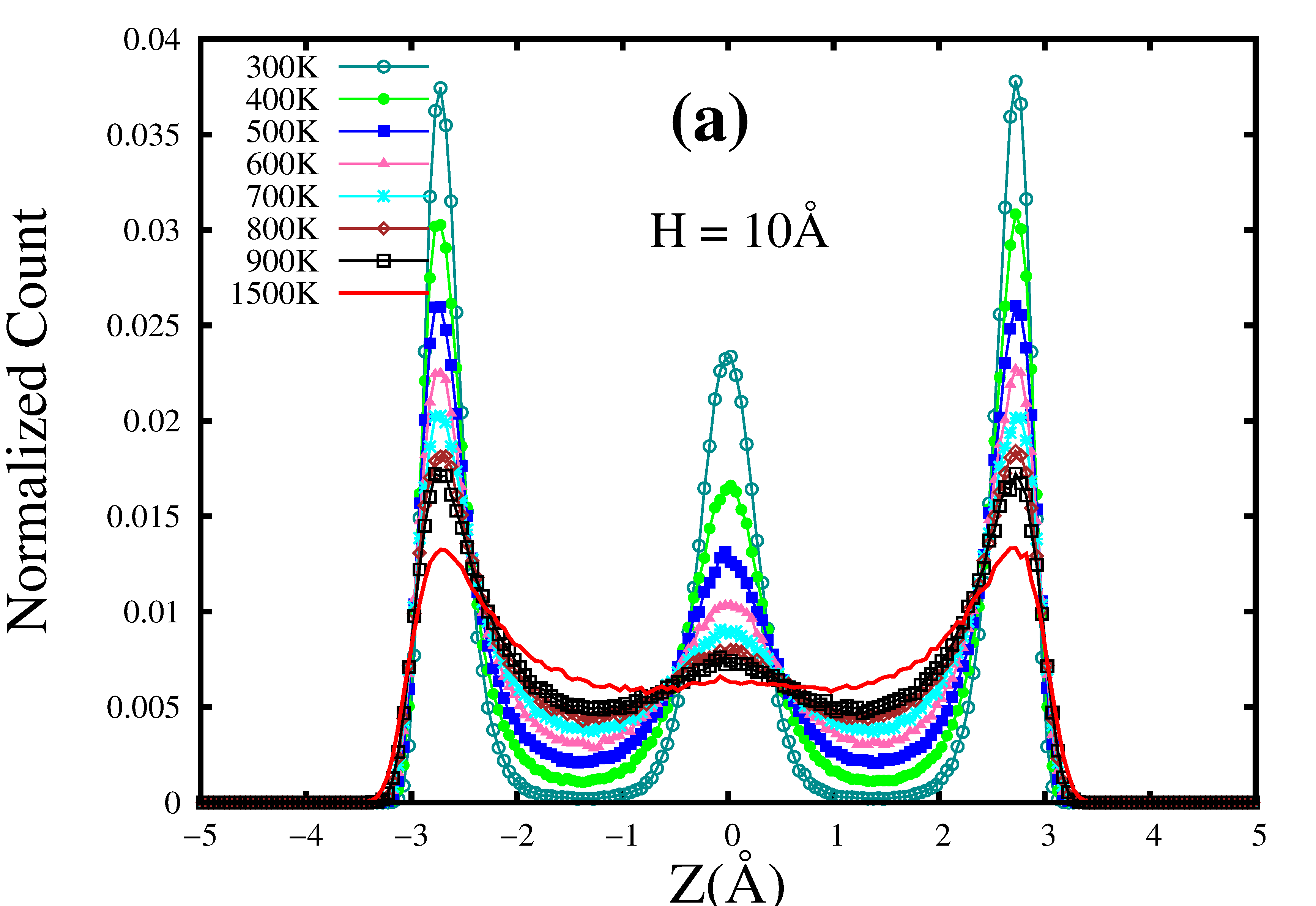}
\end{minipage}
\begin{minipage}{0.5\textwidth}
\includegraphics[width=1.05\textwidth]{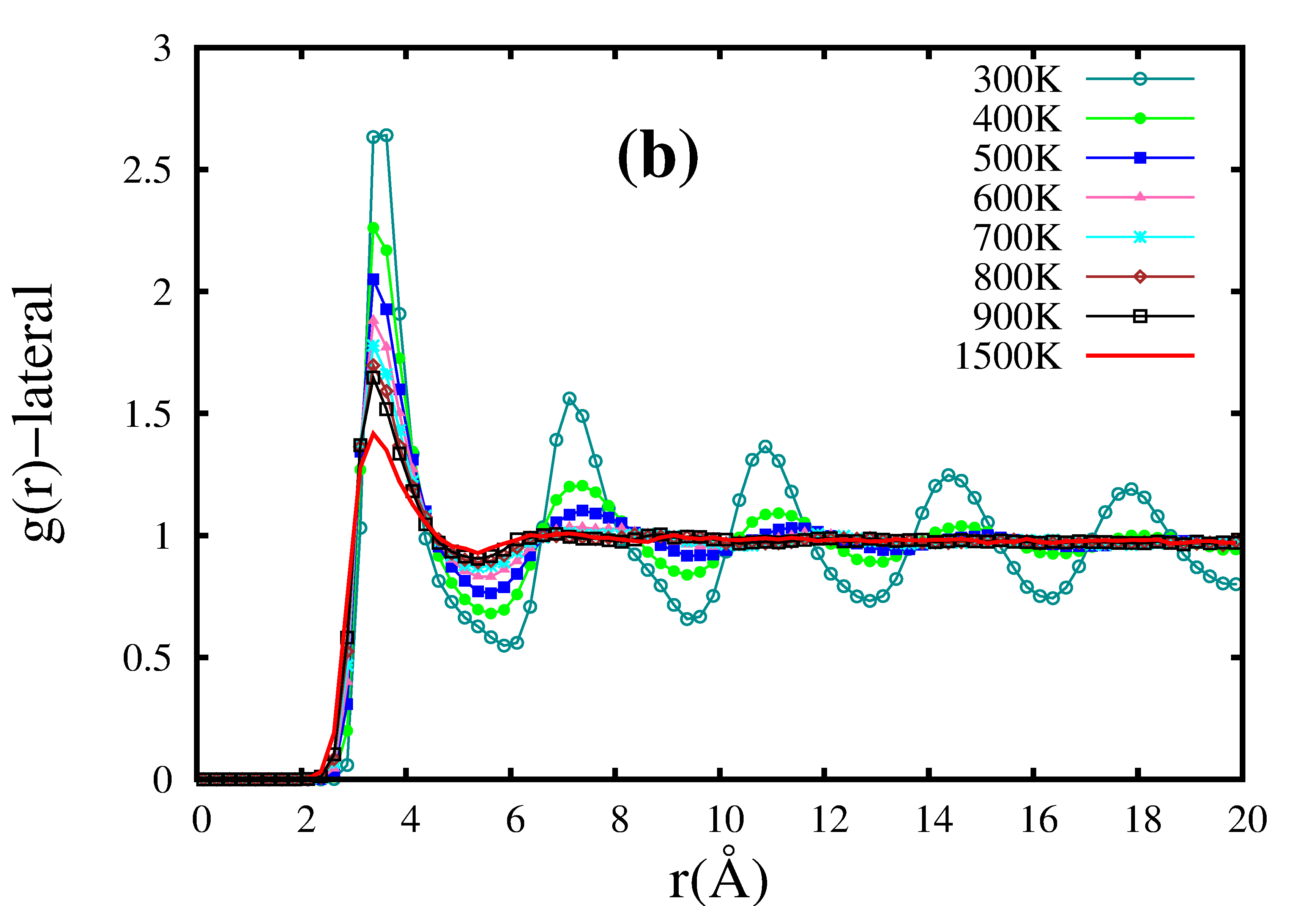}
\end{minipage}
\caption{\label{21} \textbf{(a).} Temperature evolution of the distribution of Argon particles in supercritical state with a confinement of $10$ $\AA$, averaged over several timesteps, across the Frenkel line. \textbf{(b).} $g_{\parallel}(r)$ of the central layer in the number distribution of Argon particles in supercritical state with a confinement of $10$ $\AA$, as a function of temperature, across the Frenkel line.}
\end{figure}
\end{center}
\end{widetext}
\subsubsection{\label{sec:level3}\textbf{Variation of structural features of supercritical fluid with varying rigidity of the atomistic walls:}}
In this last section we describe our investigation to analyse the role of the rigidity of the walls on the structural properties of supercritical fluid under confinement. We consider the confined spacing of $10$ $\AA$ of supercritical Argon at a state point before the Frenkel line($300$K) for this study. In our model, the wall-atoms are attached to their corresponding lattice sites by harmonic springs with the stiffness coefficient k. We vary the stiffness coefficient(k) of the springs attached to the wall-atoms from a very high value of $5000$ $ev/{\AA}^2$ (rigid, restricting the mean squared displacement (MSD) of the wall-atoms with respect to their lattice sites) to a very low value $0.005$ $ev/{\AA}^2$ (soft, increasing the mean squared displacement of the wall-atoms with respect to their lattice sites). \\
The number distribution profiles for different k values are shown in Fig.\ref{22}.(a) for supercritical Argon at $300$K with a confined spacing of $10$ $\AA$. For $10$ $ev/ \AA^2$ $\leqslant$ k $\leqslant$ $5000$ $ev/ \AA^2$, the peak-heights are found to be almost same. Further lowering of the k-value give rise to shorter peaks and for k $\leqslant$ $0.05$ $ev/ \AA^2$ the peak heights decrease considerably to give flattened distribution of particles. Further, for these k values the depletion region vanishes.\\
While the ordering is quite similar for k $>$ $0.5$ $ev/ \AA^2$, we observe a nearly $50$ $\%$ decrement of the negative pair-excess entropy ($-s^{(2)}$) for k $\leqslant$ $0.05$ $ev/ \AA^2$, which explains the flattening of peaks of the number distribution of supercritical fluid due to reduced ordering (see appendix, Table.\ref{table:3}). Decreasing the rigidity of the walls by  increasing the MSD of the wall atoms with respect to their lattice sites (lowering k value) shows a transition from an amorphous-like structure to a liquidlike ordering, parallel to the walls. Fig.\ref{22}.(b) shows the variation of $g_{\parallel}(r)$ for the central layer of H = $10$ $\AA$ spacing at $300$K. Reduction of density of the layers causes this redistribution of the particles parallel to the walls from a highly ordered amorphous phase to a comparatively less ordered liquidlike phase.     

\section{\label{sec:level1}Summary and Conclusion}
MD simulations have been carried out on bulk and partially confined supercritical LJ fluid to investigate the structural aspects of supercritical fluids across the Frenkel line. The study, done using LAMMPS, considered a system of $10^5$  particles, interacting via the Lennard-Jones potential, at P = $5000$ bar and temperatures ranging from $240$K to $1500$K simulating a wide range of densities of Argon.\\ 

VACF and RDF, evaluated using MD simulations for an isobaric line at $5000$ bar over a temperature ranging from $240$K to $1500$K, confirm the characteristics of liquidlike and gaslike phases across the Frenkel line of supercritical Argon in the bulk. In this process the Frenkel line crossover point was identified to be in the range T $\approx$ $600$-$700$K. Investigations of the density fluctuations in the bulk reveal that the changes in compressibility are consistence with the liquidlike phase going over to the gaslike phase as the Frenkel line is crossed. \\
\begin{widetext}
\begin{center}
\begin{figure}[H]
\begin{minipage}{0.5\textwidth}
\includegraphics[width=1.0\textwidth]{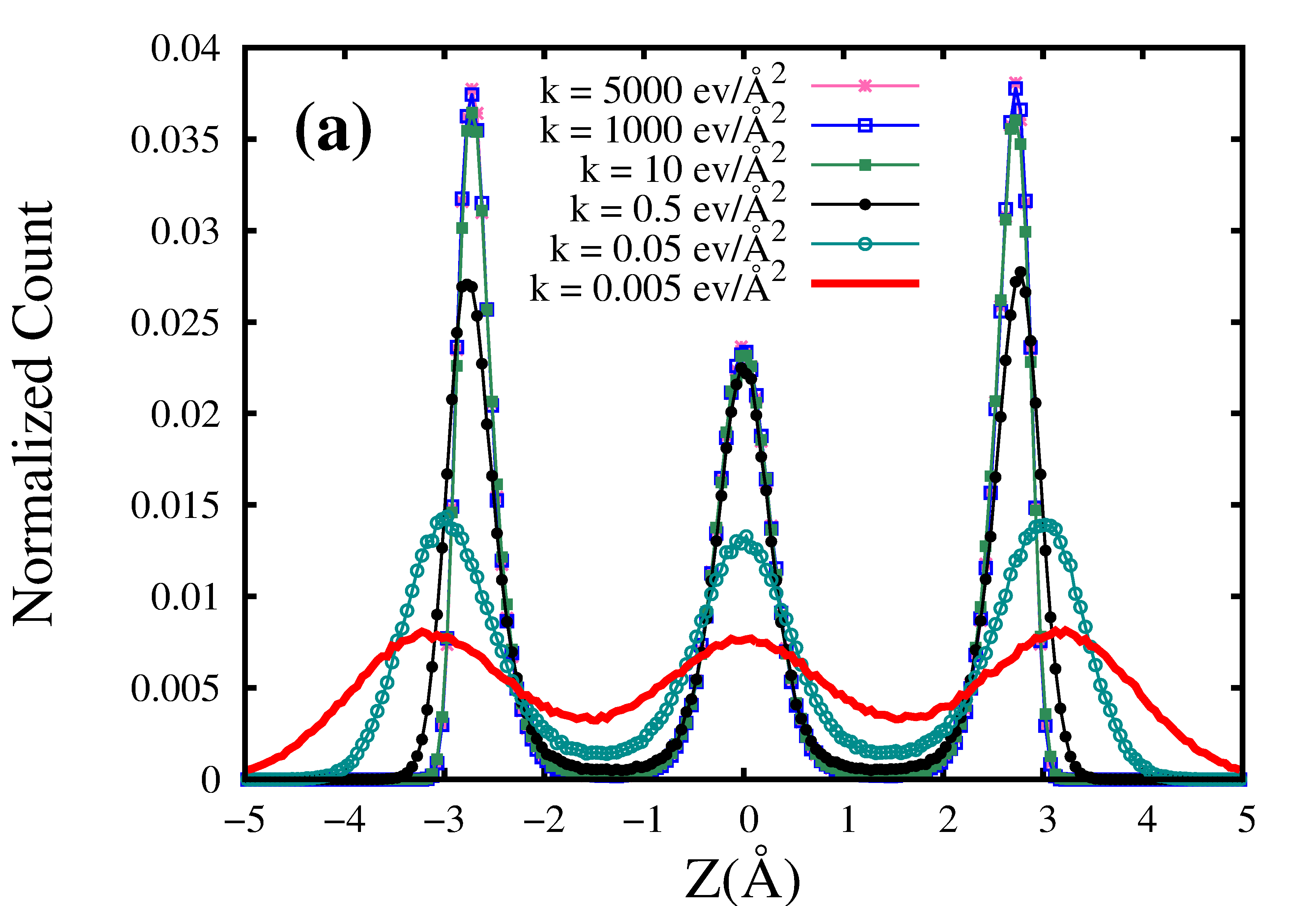}
\end{minipage}
\begin{minipage}{0.5\textwidth}
\includegraphics[width=1.0\textwidth]{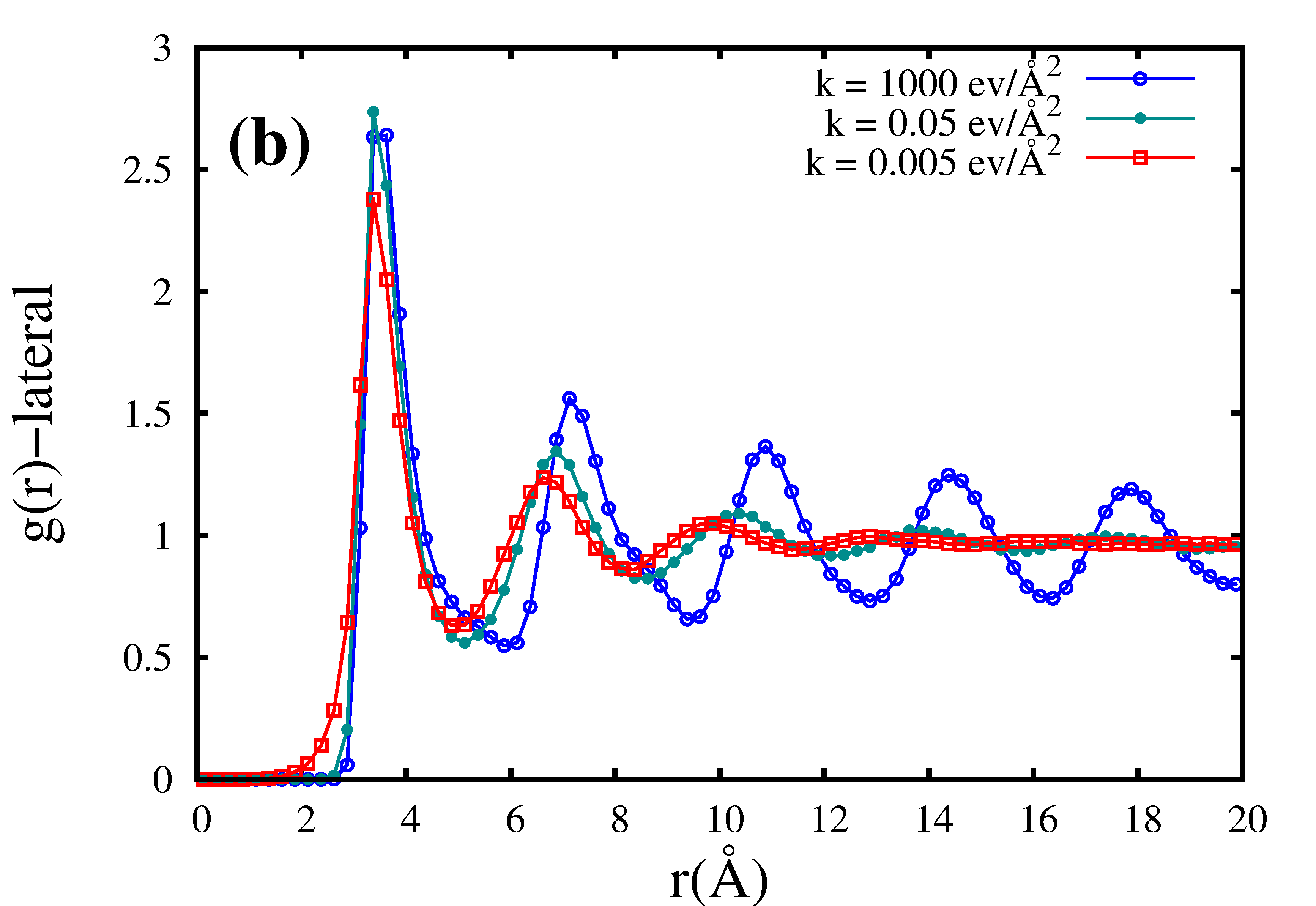}
\end{minipage}
\caption{\label{22} \textbf{(a)} Distribution of Argon particles in supercritical state, averaged over several timesteps at $300$K is calculated for different stiffness coefficients(k), normal to the walls for $10$ $\AA$ confined spacings. Different k-values are shown in different colours. \textbf{(b)} $g_{\parallel}(r)$ of the central layer in the number distribution of Argon particles in supercritical state at $300K$, with a confinement of $10$ $\AA$, as a function of stiffness coefficients (k). Lowering rigidity (lower values of k) of the wall-molecules shows transition from an amorphous to liquidlike structure parallel to the walls.}
\end{figure}
\end{center}
\end{widetext}

\onecolumngrid
\begin{widetext}
\begin{table}[H]
\centering
\caption{\label{table:1} Summary of the key features of supercritical fluids under confinement at $300$K (Before crossing Frenkel line).}
\begin{adjustbox}{max width=1.0\textwidth}
\begin{tabular}{|c|c|c|c|c|} 
\hline
& & \multicolumn{2}{|c|}{} &\\
\textbf{Confined spacing(H)($\AA$)} & \textbf{$3.4$ $\AA$ $\leqslant$ H $\leqslant$ $10$ $\AA$} & \multicolumn{2}{|c|}{\textbf{$20$ $\AA$ $\leqslant$ H $\leqslant$ $30$ $\AA$}} & \textbf{$31$ $\AA$ $\leqslant$ H $\leqslant$ $70$ $\AA$}\\
& & \multicolumn{2}{|c|}{} &\\
\hline
& & & &\\
& & \textbf{H = $20$ $\AA$, $24$ $\AA$, $30$ $\AA$} & \textbf{H = $22$ $\AA$, $26$ $\AA$, $28$ $\AA$} & \\
& & & &\\ 
\cline{3-4} 
\textbf{Distribution of particles} & Distinct peaks are observed &  & & Layers are observed \\
\textbf{normal to the walls} & with number of peaks & $\frac{H}{\sigma}$ $\approx$ integer, leading to & $\frac{H}{\sigma}$ $\not\approx$ integer, leading to & close to the walls. \\
& scaling linearly with & well-formed layers  & under-developed layers & Flat plateau develops \\
& the ratio $\frac{H}{\sigma}$ & with less packing frustration. & with more packing frustration. & around z = $0$ resembling \\
& & & & the bulk distribution.\\
& & & &\\
\hline
& & \textbf{Class P} & \textbf{Class Q} & \textbf{Class Q}\\
& & & &\\
\cline{3-5}
\textbf{Lateral RDF ($g_{\parallel}(r)$)} & Amorphous-like structures & i) Absence of peak-shift & Peak-shift in $g_{\parallel}(r)$ & Peak-shift in $g_{\parallel}(r)$\\
& are found in the layers & in $g_{\parallel}(r)$ for different layers. & for different layers & for different layers\\
& parallel to the walls. &  & are present. & are present.\\
&  & ii) H = $30$ $\AA$ shows &  & \\
&  & amorphous-like structure. &  & \\
\hline
\end{tabular}
\end{adjustbox}
\end{table}
\end{widetext}
For the first time, MD simulations of confined supercritical fluid are reported for two P-T state points, one before (P= $5000$ bar, T=$300$K) and one after crossing the Frenkel line of supercritical Argon (P= $5000$bar, T=$1500$K) using both smooth and atomistic walls. At each P-T state point, the confinement spacing ranged from very narrow spacings like $3.4$ $\AA$ (= $1$ atomic diameter) to larger spacing of $70$ $\AA$ ($\approx$ $21$ atomic diameters) while maintaining a constant density corresponding to the chosen P-T state. We further investigate the effect of confinement in the vicinity of the Frenkel line by considering state points across the line. The effect of rigidity of the walls on the structural properties of supercritical fluids is also studied in the context of confinement. \\ 
In the "liquidlike" regime (point A, before crossing the Frenkel line), layering of particles perpendicular to the confining walls is pronounced. Analysis of the component of RDF that is parallel to the confining walls of successive layers reveals that the particles arrange themselves in a close-packing formation when smooth walls are imposed as boundaries(see appendix, Fig.\ref{27}). On the contrary, in the presence of atomistic walls this close-packing formation breaks down due to the appearance of near-wall depletion layers and associated packing frustration in the confinement. Also more ordered patterns are observed under atomistic wall-confined systems. The accommodation of the particles under confinement is governed by spacing between the walls, which, depending on less or more frustration in packing, selectively allows particles to form well-developed layer along $z$. We have found a spacing(H = $30$ $\AA$) at $300$K (before crossing Frenkel line), where the distribution of particles exhibit maximum ordering and amorphous-like structure forms parallel to the walls. Immediately after which ordering disappears around $z$ = $0$. Rigidity of atomistic walls play a crucial role on the structural properties of the supercritical fluids. We observe a significant loss of ordering of the fluid, both normal and parallel to the walls, on modelling the walls softer by increasing the average MSD of the wall-atoms with respect to its lattice sites. This correlation between the ordering of fluids and the rigidity of the walls, has been confirmed by the two-body excess entropy measurements.\\

Extreme confinements with very narrow widths (spacing $\sim$ $1$, $2$ or $3$ times the diameter of Argon) show prominent layering which can be enhanced using higher interaction strength between wall and fluid. More interestingly, amorphous-like structural features are confirmed by $g_{\parallel}(r)$ along these well-defined layers parallel to the walls. Further, studying state points at the vicinity of the Frenkel line under confinement shows the gradual disappearance of layering across the Frenkel line. In the “gaslike” regime (point B, after crossing the Frenkel line), such ordering is not seen apart from the region very close to the walls.\\

The structural ordering has been quantified by two-body excess entropy and translational order parameter calculations which clearly indicate the correlation between confinement and ordering in a supercritical fluid. 
\begin{widetext}
\begin{figure}[H]
\centering
\includegraphics[width=1.0\textwidth]{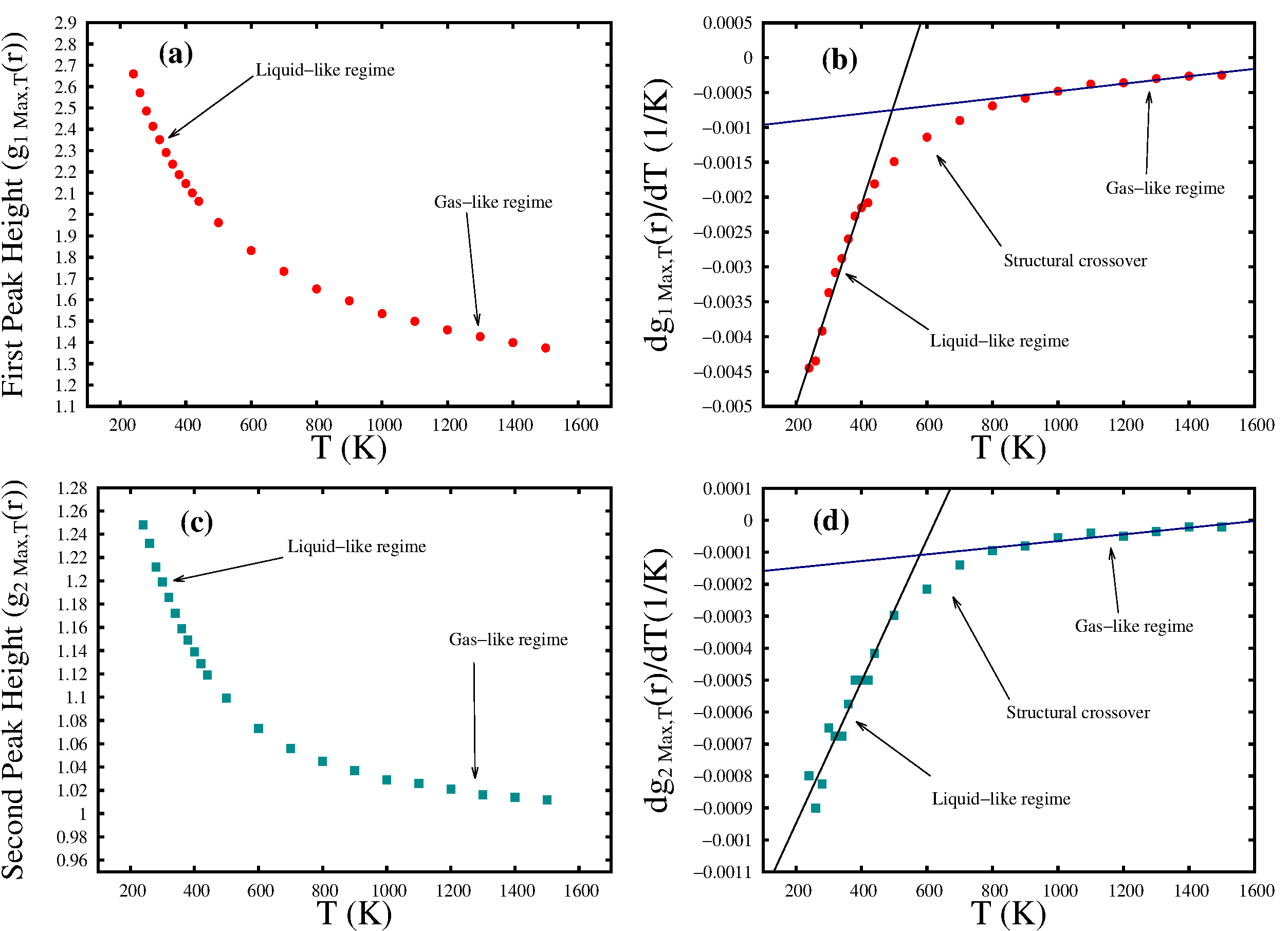}
\caption{\label{23} \textbf{(a)} First peak-height of RDF as a function of temperature (T). \textbf{(b)} Variation of the First order temperature derivative of the first peak-height of RDF with temperature (T). \textbf{(c)} Second peak-height of RDF as a function of temperature (T). \textbf{(d)} Variation of the First order temperature derivative of the second peak-height of RDF with temperature (T).}
\end{figure}
\end{widetext}
The discrete jump in translational order parameter at the spacing (H = $30$ $\AA$) confirms the sudden loss of ordering for H $>$ $30$ $\AA$ (Fig.\ref{12}.(a)).\\
While layering is known to occur in normal fluids at high density, it is not typical of liquids to show amorphous-like structures parallel to the walls under confinement at $300$K. Close-packing like structural feature, found in supercritical fluids in the liquidlike regime under smooth walls, is also not a  general feature for liquids. Since layering is not significant in the "gaslike" phase of the supercritical fluid, it is unlikely to occur for normal fluids in the "gaslike" regimes as they have much lower densities.\\ 
The structural aspects of very narrow confined systems further open up possibilities of unusual dynamics parallel to the walls, where "liquidlike" phase of SCF suffers a 
\begin{table}[H]
\centering
\caption{\label{table:2} Standard Deviation for the density fluctuations and isothermal compressibility ($\kappa_T$) for bulk supercritical Argon like LJ fluid for  different temperatures at $5000$ bar along the Frenkel line.}
\begin{tabular}{|p{0.7cm}||p{0.7cm}||p{3.6cm}||p{2.6cm}|} 
\hline
& & &\\
\textbf{P} & \textbf{T} & \textbf{$\sigma$(Standard Deviation} & \textbf{$\kappa_T$
($10^{-10}$ $Pa^{-1}$)}\\ [0.5ex]
\textbf{(bar)} & \textbf{(K)} & \textbf{in Normalized unit)} & \\ [0.5ex]
\hline
& & &\\
& \hspace{0.2cm}$240$ & $\hspace{0.5cm}0.58\times 10^{-3}$  & $\hspace{0.5cm} 4.189$ \\
& \hspace{0.2cm}$300$ & $\hspace{0.5cm}0.67\times 10^{-3}$ & $\hspace{0.5cm} 4.717$ \\
& \hspace{0.2cm}$500$ & $\hspace{0.5cm}0.91\times 10^{-3}$ & $\hspace{0.5cm} 6.091$ \\
\hspace{0.2cm}$5000$ & \hspace{0.2cm}$700$ & $\hspace{0.5cm}1.11\times 10^{-3}$ & $\hspace{0.5cm} 7.219$ \\
& \hspace{0.2cm}$900$ & $\hspace{0.5cm}1.28\times 10^{-3}$ & $\hspace{0.5cm} 8.285$ \\
& \hspace{0.2cm}$1000$ & $\hspace{0.5cm}1.34\times 10^{-3}$ & $\hspace{0.5cm} 8.568$ \\
& \hspace{0.2cm}$1500$ & $\hspace{0.5cm}1.71\times 10^{-3}$ & $\hspace{0.5cm} 11.393$ \\[1ex] 
\hline
\end{tabular}
\end{table}

\onecolumngrid
\begin{widetext}
\begin{figure}[H]
\centering
\includegraphics[width=1.0\textwidth]{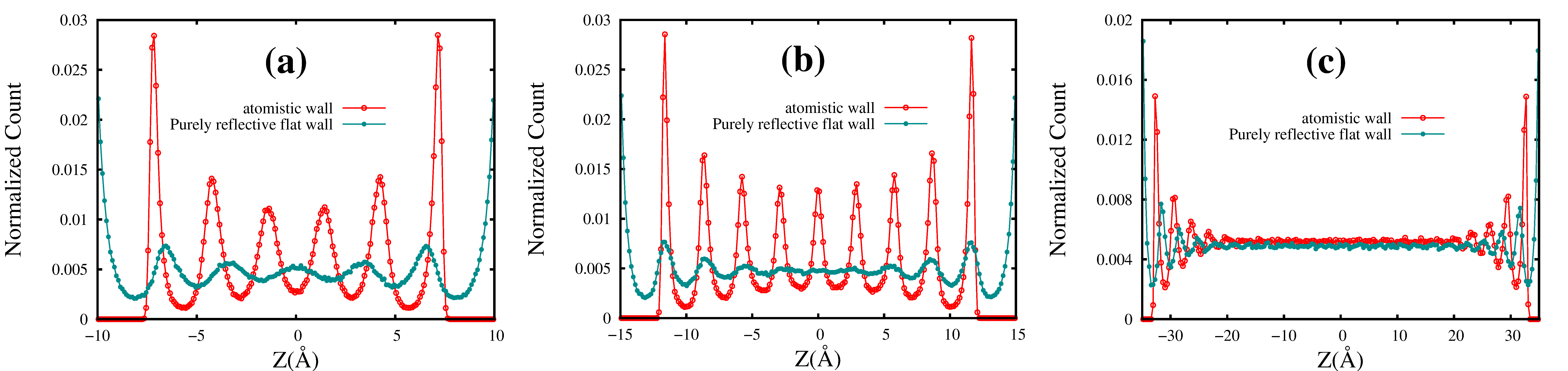}
\caption{\label{24} Distribution of Argon particles in supercritical state, averaged over several timesteps at $300$K normal to the walls for different confined spacings(H) \textbf{(a)}H=$20$ $\AA$, \textbf{(b)}H=$30$ $\AA$, \textbf{(c)}H=$70$ $\AA$.}
\end{figure}
\end{widetext}
transition to amorphous-like fluid. Also, our findings on the wall-rigidity dependence of the structural features of SCF offer new scopes to elucidate the relation between the supercritical fluid and wall dynamics. The key features of our findings are summarized in a tabular form in Table \ref{table:1}.\\
The presence of ordering in the ‘liquidlike” regime and the absence of it in the “gaslike” regime under confinement may have significant implications for the variation of transport properties across the Frenkel line. This heterogeneity, present both in bulk and confined systems, might be responsible for breakdown of the universal scaling  between structure and dynamics of fluids and stimulates possibilities of having a unique relationship between them. 
\section{\label{sec:level1}Acknowledgement}
We acknowledge the help of HPCE,IIT Madras for high performance computing. KG would like to express his gratitude to Department of Science and Technology(DST)(India) for INSPIRE Fellowship for funding.

\appendix

\section{Bulk studies of supercritical fluid}
\subsection{\label{sec:level1}Temperature variation of First and second peak-height of RDF(radial distribution function) and its derivatives in bulk phase of supercritical Argon for all P,T state points chosen for study at P = 5000 bar: Identification of Frenkel line:}
Figure \ref{23} describes the temperature variation of the first
and second RDF peaks in detail.
\begin{widetext}
\begin{center}
\begin{figure}[H]
\begin{minipage}{.5\textwidth}
\includegraphics[width=0.95\textwidth]{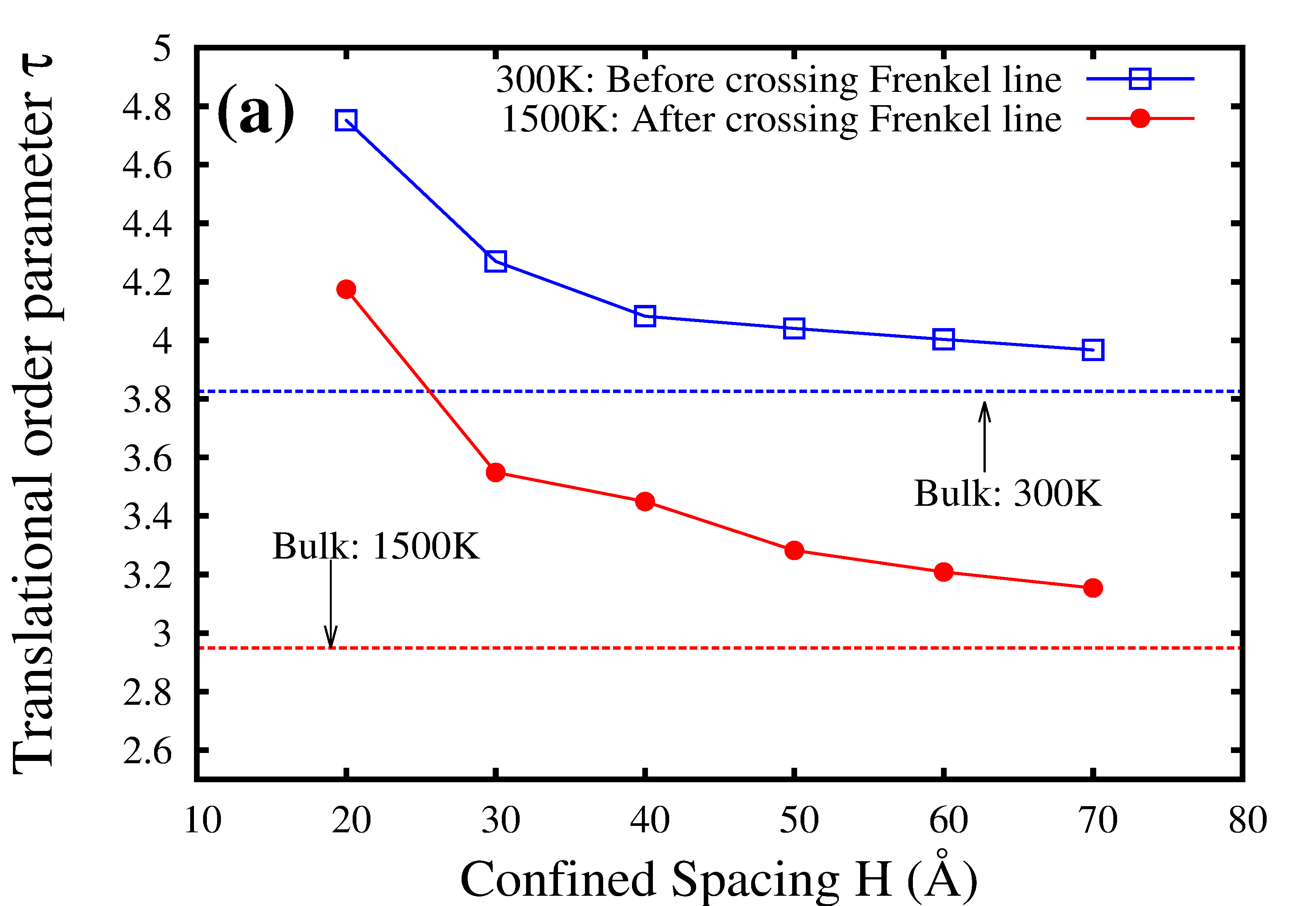}
\end{minipage}
\begin{minipage}{.5\textwidth}
\includegraphics[width=0.95\textwidth]{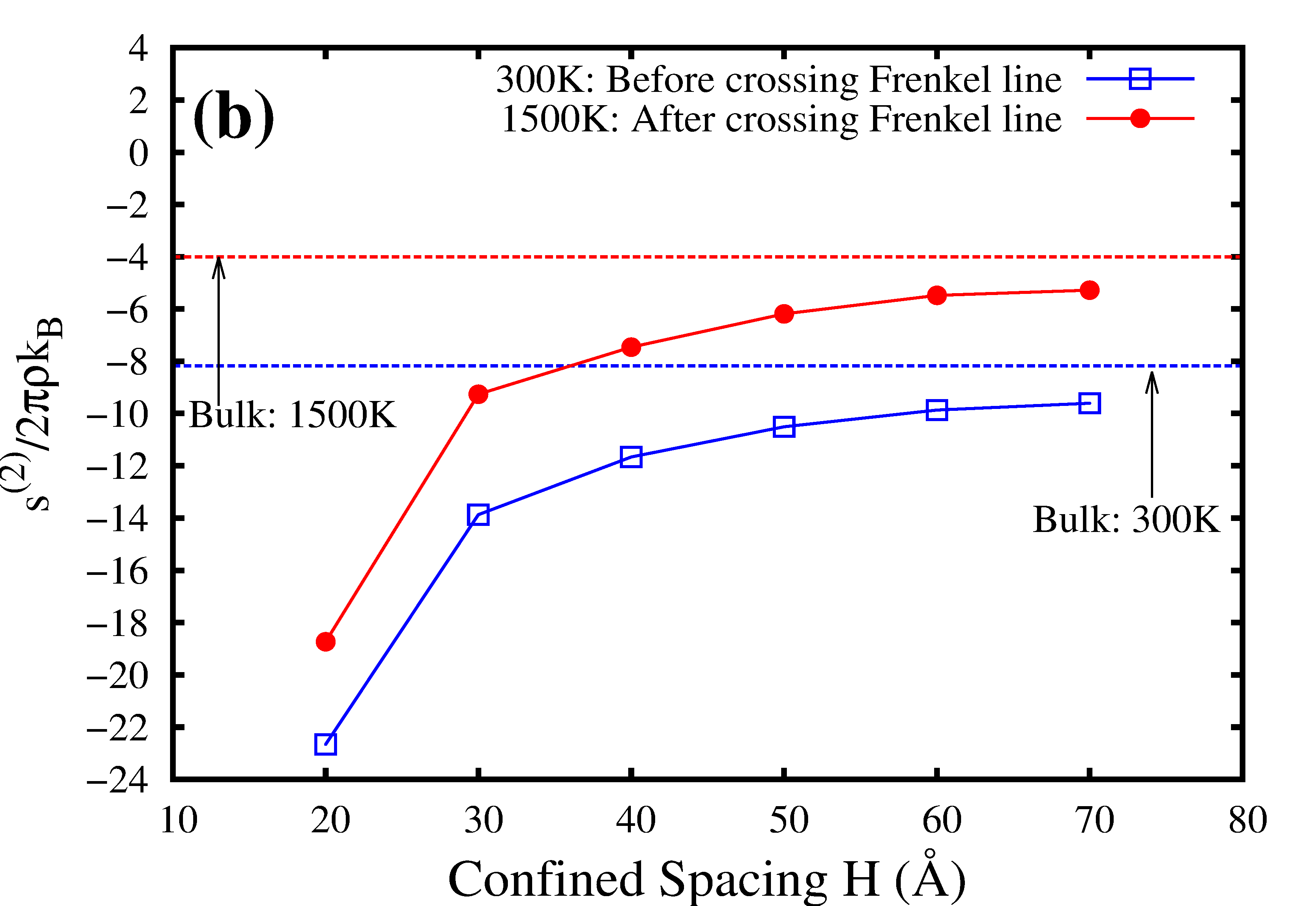}
\end{minipage}
\caption{\label{25} \textbf{(a).} Translational order parameter ($\tau$) variation with different confined spacings with smooth walls both before and after crossing the Frenkel line of supercritical Argon. For reference the bulk phase values have also been shown. \textbf{(b).} Scaled two-body excess entropy ($s^{(2)}/2 \pi \rho k_B$) variation with smooth walls for different confined spacings both before and after Frenkel line of supercritical Argon. The corresponding Bulk phase values have also been shown for reference .} 
\end{figure}
\end{center}
\end{widetext}
\onecolumngrid
\begin{widetext}
\begin{center}
\begin{figure}[H]
\begin{minipage}{.5\textwidth}
\includegraphics[width=0.95\textwidth]{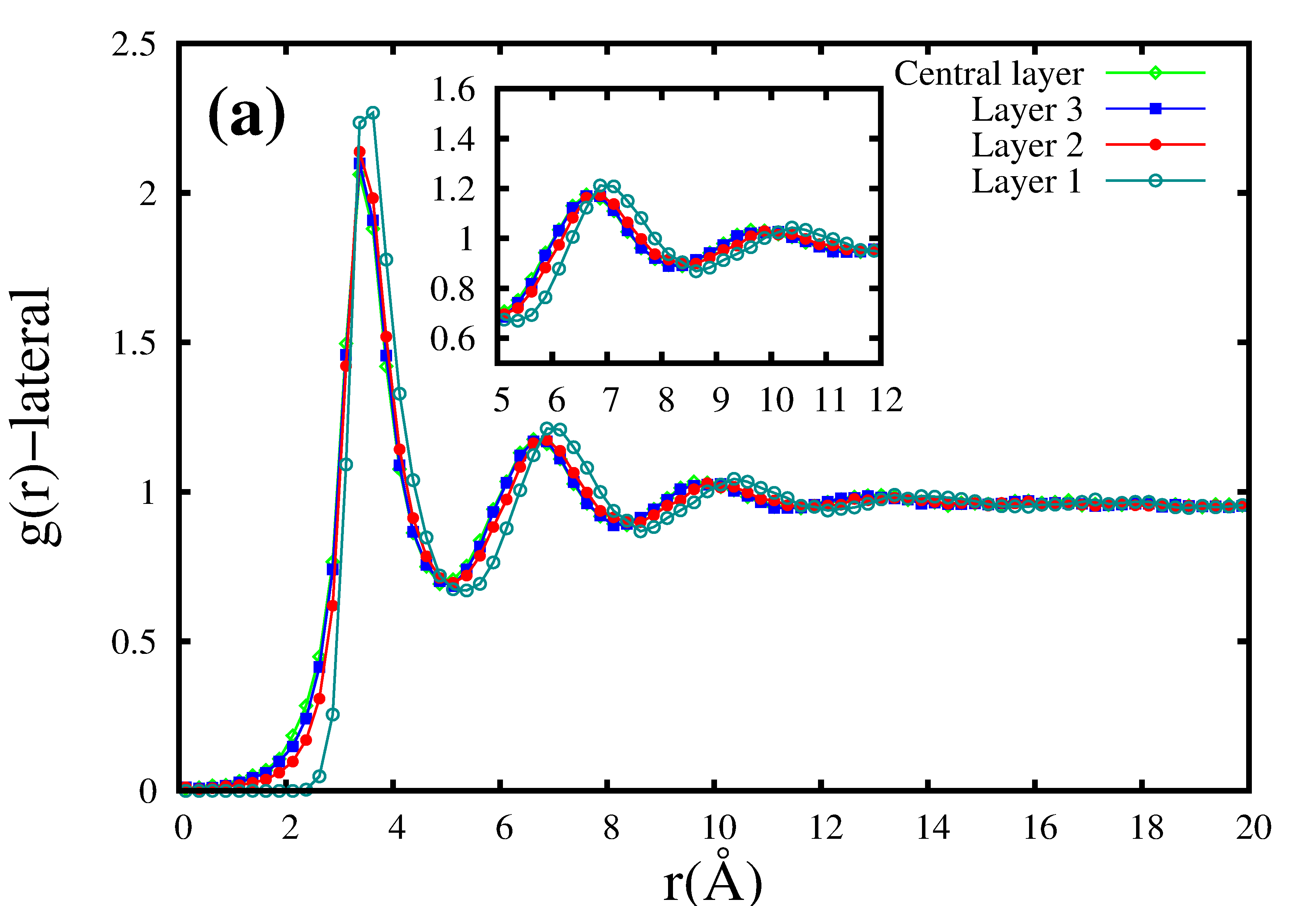}
\end{minipage}
\begin{minipage}{.5\textwidth}
\includegraphics[width=0.95\textwidth]{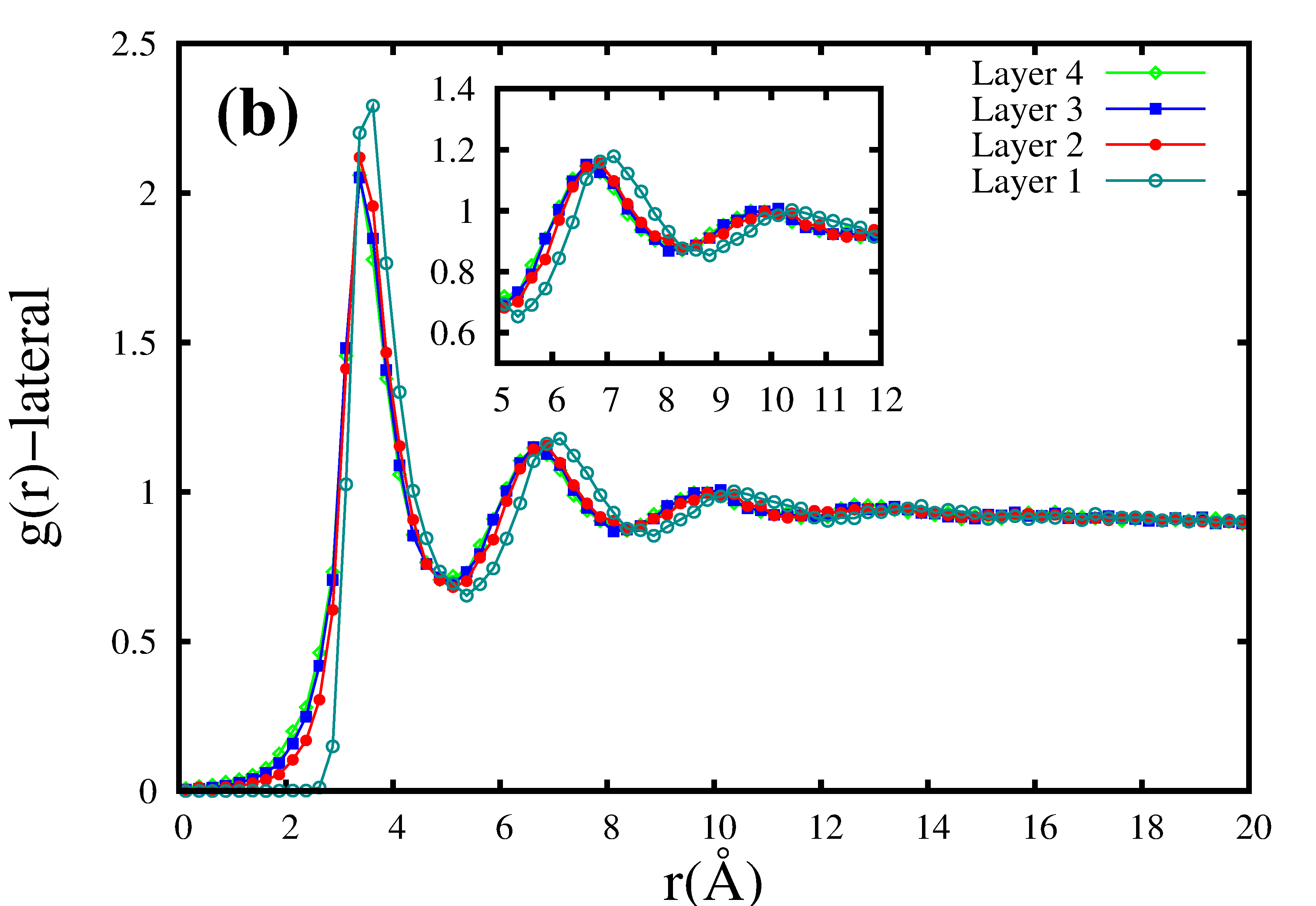}
\end{minipage}
\vspace{1cm}
\caption{\label{26} Radial Distribution function ($g_{\parallel} (r)$ of Argon particles in supercritical phase under different confined spacings (H) at $300$K parallel to the smooth, purely reflective walls.\textbf{(a)}H=$20$ $\AA$ and \textbf{(b)}H=$70$ $\AA$. Insets provide the details of the second and third RDF peaks parallel to the smooth walls. By symmetry, only one set of layers with respect to the centre is considered. Numbering of layers starts from the layer closest to the wall. }
\end{figure}
\end{center}
\end{widetext}
\section{Dataset for Number Density Fluctuations and Compressibility in bulk supercritical Argon}
Table. \ref{table:2} shows the dataset of compressibility in bulk
supercritical argon at 5000 bar across the Frenkel line.
\section{Confinement studies of supercritical fluid}
\subsection{\label{sec:level2} Layering of supercritical fluid normal to the walls before crossing the Frenkel line(300K): Comparison between smooth, purely reflective and atomistic walls}
We observe at wider spacings (e.g. H = $70$ $\AA$) the number distribution profiles are quite similar with the presence of bulk number distribution around $z$= $0$ for both flat and atomistic walls. Prominent differences can be seen as we go towards smaller spacings (H = $20$ $\AA$ and $30$ $\AA$) (Fig.\ref{24}).\\
\subsection{\label{sec:level2} Translational order parameter and Two-body excess entropy studies normal to the purely reflective walls:}
The smooth, purely reflective boundaries manifest nearly monotonous trends of $\tau$ and $s^{(2)}$ as a function of confined separations. $\tau$ and $s^{(2)}$ show monotonically decreasing and increasing trends respectively towards the bulk value with increasing spacing (Fig.\ref{25}) . 

\subsection{\label{sec:level2} Radial distribution function parallel to the smooth, purely reflective walls before crossing Frenkel line:}

\begin{figure}[H]
\centering
\includegraphics[width=0.5\textwidth]{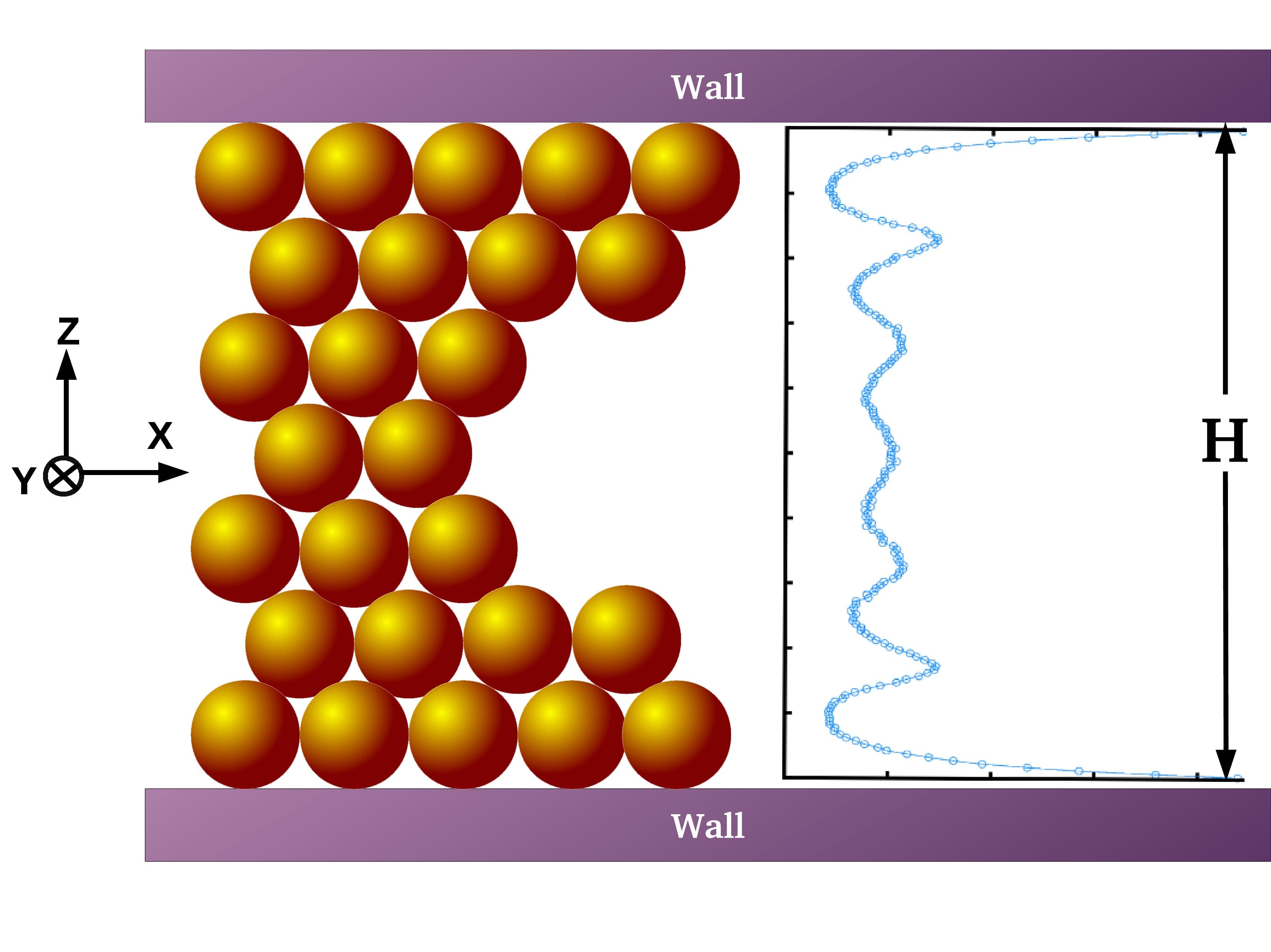}
\caption{\label{27} The schematic diagram of the close packing structure of supercritical Argon before crossing the Frenkel line at $5000$ bar pressure for confined system with small spacing(H = $20$ $\AA$) under smooth, purely reflective walls.}
\end{figure}

Smooth, purely reflective boundaries, unlike to the atomistic walls, show only one class of $g_{\parallel}(r)$ where positional shift of the coordination spheres is found for all spacings ranging from $20$ $\AA$ to $70$ $\AA$ (class Q) (Fig.\ref{26}). These positional shifts, along with the periodic the mono-layer formation along $z$, suggest a close packing structure parallel to the flat walls and are depicted in Fig.\ref{27}. For small spacings the close packing structure is seen to prevail across the entire thickness (Fig.\ref{27}). 
As the spacing(H) between the walls increases, the close packing arrangement gradually seems to disappear in the central region of the confinement.
\subsection{\label{sec:level2} Radial distribution function parallel to the smooth, purely reflective walls after crossing Frenkel line:}
\begin{figure}[H]
\centering
\includegraphics[width=0.5\textwidth]{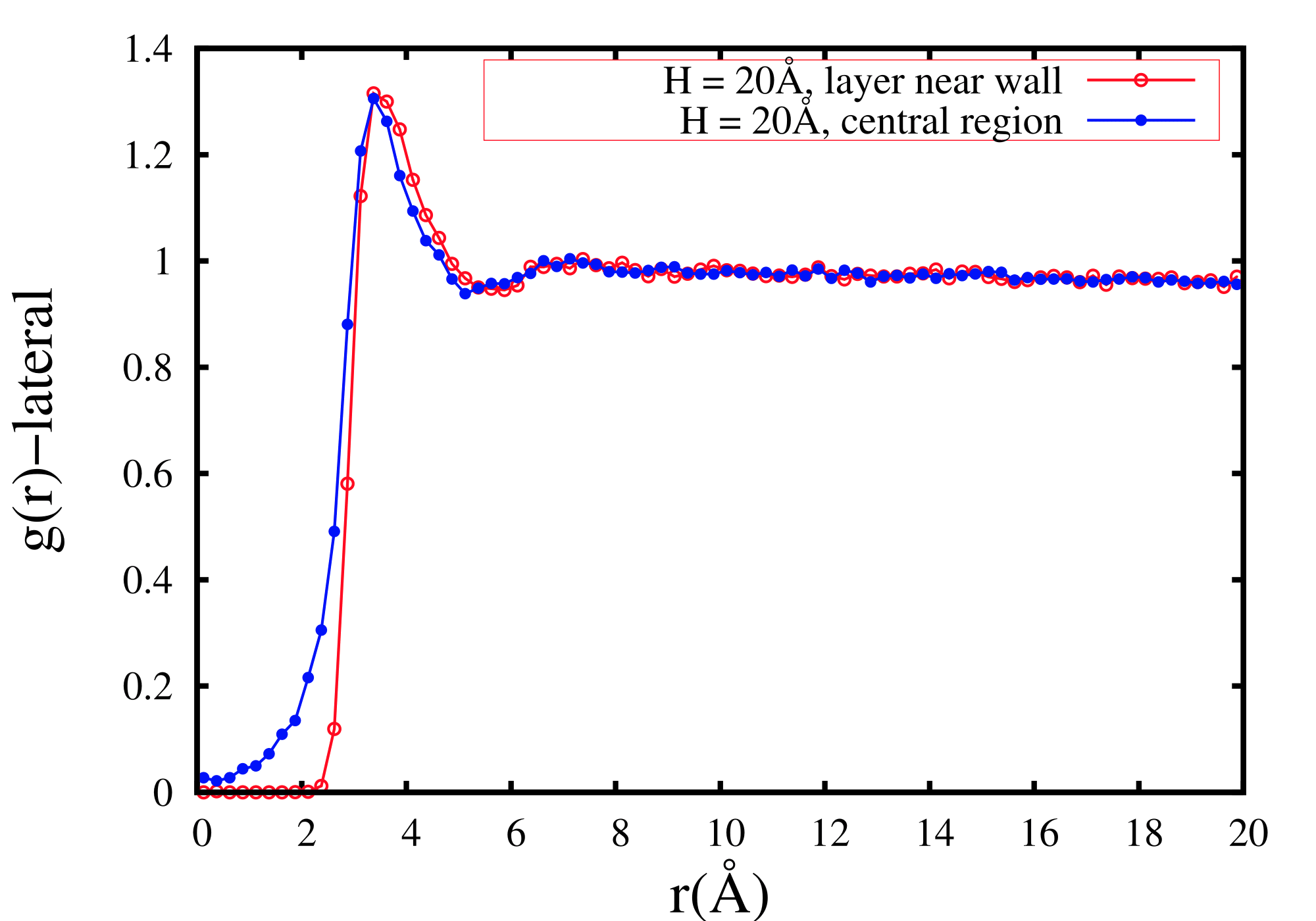}
\caption{\label{28} The nature of lateral RDF ($g_{\parallel}(r)$)of central region and region close to the smooth, purely reflective walls after crossing Frenkel line at $1500$K temperature.}
\end{figure}

After crossing the Frenkel line, layering becomes insignificant. Fig.\ref{28} shows one such case at $1500$K temperature for $20$ $\AA$ wall-spacing for smooth walls, where we don't see much change in comparing the $g_{\parallel}(r)$ of central region and that of the region close to the walls. 
\begin{widetext}
\subsection{\label{sec:level2} Excess entropy dataset for different k values (stiffness coefficient)of the walls for $10$ $\AA$ spacing at $300$K:}
Table \ref{table:3} describes the variation of pair-excess entropy as a function of stiffness co-efficients of the walls.
\begin{table}[H]
\caption{\label{table:3} Pair-excess entropy of Argon particles in supercritical phase for H = $10$ $\AA$ at $300$K as a function of stiffness coefficient (k). A nearly $50$ $\%$ decrement of the negative pair-excess entropy ($-s^{(2)}$) for k $\leqslant$ $0.05$ $ev/ \AA^2$ has been observed, while the ordering is found to be similar for k $>$ $0.5$ $ev/ \AA^2$.}
\centering
\begin{tabular}{|p{2.0cm}||p{6.0cm}|} 
\hline
& \\
\hspace{0.4cm}\textbf{k($ev/{\AA}^2$)} & \hspace{0.4cm} \textbf{$-s^{(2)}/2\pi\rho k_{B}$ (pair-excess entropy)} \\ [0.5ex]
\hline
& \\
\hspace{0.9cm}$5000$ & \hspace{0.9cm}$15.51$ \\
\hspace{0.9cm}$1000$ & \hspace{0.9cm}$16.25$\\
\hspace{0.9cm}$10$ & \hspace{0.9cm}$15.85$ \\
\hspace{0.9cm}$0.5$ & \hspace{0.9cm}$10.23$\\
\hspace{0.9cm}$0.05$ & \hspace{0.9cm}$7.95$ \\
\hspace{0.9cm}$0.005$ & \hspace{0.9cm}$6.55$ \\[1ex] 
\hline
\end{tabular}
\end{table}
\end{widetext}

\providecommand{\noopsort}[1]{}\providecommand{\singleletter}[1]{#1}%
\end{document}